\documentclass[11pt,letterpaper]{article}

\usepackage{lmodern}
\usepackage{latexsym,multirow}
\usepackage{amssymb,amsmath,bm}
\usepackage{bigints}
\usepackage{array}
\usepackage{longtable}
\usepackage{booktabs}       
\usepackage{bbm} 
\usepackage{graphicx,float}
\usepackage[color,all,import,arrow]{xy}
\usepackage{enumerate}
\usepackage{verbatim}
\usepackage{lscape}
\usepackage{mathtools}
\usepackage{fancyvrb}

\usepackage[format=plain,
            labelfont={bf,it},
            textfont=it]{caption}

\usepackage{longtable}

\usepackage{placeins}

\usepackage[natbib=true,minnames=1,maxnames=2,backend=biber,style=authoryear-luh-ipw]{biblatex} 
\usepackage{atbegshi}
\AtBeginDocument{\AtBeginShipoutNext{\AtBeginShipoutDiscard}}

\usepackage{dcolumn}
\newcolumntype{.}{D{.}{.}{-1}}
\newcolumntype{d}[1]{D{.}{.}{#1}}

\graphicspath{{.}}

\usepackage{caption}
\usepackage{subcaption}
\DeclareCaptionLabelFormat{step}{Step \textbf{#2)}} 

\usepackage[T1]{fontenc}
\usepackage{tikz}
\usepackage{qtree}
\usepackage{comment}
\usetikzlibrary{arrows}
\usetikzlibrary{positioning}
\usetikzlibrary{matrix}
\usetikzlibrary{bayesnet}
\usetikzlibrary{decorations.pathreplacing,calligraphy} 

\usepackage[ruled,vlined]{algorithm2e}

\usepackage{theorem}
\theoremstyle{plain}
\theoremheaderfont{\scshape}

\usepackage{chngpage}

\newcommand{\RATEWT}{AUTOC} \newcommand{\RATEWTALT}{QINI} 

\usepackage{rotating}

\usepackage{arydshln}

\usepackage[compact]{titlesec}

\usepackage{setspace}
\usepackage[bottom]{footmisc}
\allowdisplaybreaks

\newcommand\spacingset[1]{\renewcommand{\baselinestretch}%
{#1}\small\normalsize}

\newcommand{\blind}{0}

\newcommand{\E}{\mathbb{E}}
\newcommand{\bX}{\mathbf{X}}
\newcommand{\bx}{\mathbf{x}}

\newcommand{\bM}{\mathbf{M}}

\newcommand{\bV}{\mathbf{V}}

\newcommand{\bv}{\mathbf{v}}

\usepackage[pdftex, bookmarksopen=true, bookmarksnumbered=true,
pdfstartview=FitH, breaklinks=true,
urlbordercolor={0 1 0}, citebordercolor={0 0 1}]{hyperref}

\usepackage{xr}

\begin{document} 

\newcommand{\tit}{
Effect Heterogeneity with Earth Observation in Randomized Controlled Trials: Exploring the Role of Data, Model, and Evaluation Metric Choice
}
%
\spacingset{1.25}

\if0\blind

{\title{\bf\tit\thanks{
Code implementing the paper analyses can be found at \href{https://github.com/AIandGlobalDevelopmentLab/causalimages-software}{\url{GitHub.com/AIandGlobalDevelopmentLab/causalimages-software}}. 
} 
  \author{Connor T.\, Jerzak
\thanks{Assistant Professor, Department of Government, University of Texas at Austin. ORCID: 0000-0003-1914-8905  Email: \href{mailto:connor.jerzak@austin.utexas.edu}{connor.jerzak@austin.utexas.edu} URL:
      \href{https://connorjerzak.com}{\texttt{ConnorJerzak.com}}
      }
   \and 
   Ritwik Vashistha\thanks{PhD Student, Department of Statistics \& Data Science, University of Texas at Austin. ORCID: 0000-0003-1914-8905. Email: \href{mailto: ritwik.v@utmail.utexas.edu}{ritwik.v@utmail.utexas.edu} URL:
      \href{https://ritwik.super.site}{\texttt{ritwik.super.site}}
      }
   \and 
   Adel Daoud\thanks{Stanford University \& Link\"{o}ping University. ORCID: 0000-0003-1914-8905. Email: \href{mailto:adel.daoud@liu.se}{adel.daoud@liu.se} URL:
      \href{https://AdelDaoud.se}{\texttt{AdelDaoud.se}} AI \& Global Development Lab: \href{https://ritwik.super.site}{\texttt{global-lab.ai}}
      \href{https://AdelDaoud.se}{AdelDaoud.se} AI \& Global Development Lab: \href{https://global-lab.ai}{\texttt{global-lab.ai}}
      }
    }
  \date{
    \today
  }
}
}

\fi

\if1\blind
\title{\bf \tit}
\fi

\maketitle

\pdfbookmark[1]{Title Page}{Title Page}

\thispagestyle{empty}
\setcounter{page}{0}

\vspace{-1cm}
\begin{abstract}
  \noindent Many social and environmental phenomena are associated with macroscopic changes in the built environment, captured by satellite imagery on a global scale and with daily temporal resolution. While widely used for prediction, these images and especially image sequences remain underutilized for causal inference, especially in the context of randomized controlled trials (RCTs), where causal identification is established by design. In this paper, we develop and compare a set of general tools for analyzing Conditional Average Treatment Effects (CATEs) from temporal satellite data that can be applied to any RCT where geographical identifiers are available. Through a simulation study, we analyze different modeling strategies for estimating CATE in sequences of satellite images. We find that image sequence representation models with more parameters generally yield a greater ability to detect heterogeneity. To explore the role of model and data choice in practice, we apply the approaches to two influential RCTs---\citet{banerjee2015}, a poverty study in Cusco, Peru, and \citet{bolsen2014voters}, a water conservation experiment in Georgia, USA. We benchmark our image sequence models against image-only, tabular-only, and combined image-tabular data sources, summarizing practical implications for investigators in a multivariate analysis. Land cover classifications over satellite images facilitate interpretation of what image features drive heterogeneity. We also show robustness to data and model choice of satellite-based generalization of the RCT results to larger geographical areas outside the original. Overall, this paper shows how satellite sequence data can be incorporated into the analysis of RCTs, and provides evidence about the implications of data, model, and evaluation metric choice for causal analysis. 
\vspace{.00cm}
\\ \noindent {\bf Keywords: } Heterogeneous treatment effects; Earth observation; Randomized controlled trials; Transportability 
\end{abstract}

\spacingset{1}
\section*{Introduction}
Although randomized control trials (RCT) deployed in complex social settings tend to produce unbiased Average Treatment Effect (ATE) estimates, they are costly to run. The more covariates required, the more expensive the trial. An RCT trial deployed in Africa may cost several hundred thousand to millions of USD, depending on the trial size, duration, and location of the experimental subjects. The cost of RCTs may limit researchers' ability to analyze variability in treatment response--through Conditional Average Treatment Effects (CATEs)---because researchers will often confine CATE analysis to only using variables reported by experimental subjects---thus, only available for those units from around the period under study (i.e., in variables such as respondents' current income). Consequently, historical and contextual information may be underplayed in contemporary CATE analyses.

Herein lies the potential of using satellite images \citep{burkeUsingSatelliteImagery2021,kluger2022combining,daoud_statistical_2023}. These images provide a cost-effective way to complement traditional tabular data when gathering information about large areas or extended time periods. This cost-effectiveness becomes especially significant in large-scale RCT studies. A satellite image captures the location of the subjects, measuring the visual appearances of their living conditions, the spatial arrangement of their neighborhoods, and accessibility to available resources. A sequence of satellite images measures changes in these appearances and arrangements---capturing historical processes. Accordingly, sequences of satellite imagery in RCTs are likely to enhance the robustness and scope of research across various disciplines.

To see why, we first note that satellite sequence data may provide enhanced environmental contextualization. These images provide comprehensive environmental information relevant to poverty, agriculture, ecology, and epidemiology \citep{kinoScopingReviewUse2021,sakamoto_planetary_2024}. Image sequence data may help researchers account for external variables such as climate, land use, and natural resources, all of which might influence RCT outcomes. 

Second, satellite images supply scalability and coverage. Such imagery allows for the examination of large geographical areas without the need for extensive on-ground data collection. This capability is invaluable for studies that require data from remote or inaccessible regions, enabling more inclusive data analyses. Also, satellite data provide an objective measure that is broadly consistent across different locations and times. This uniformity is crucial for the comparability of the underlying data across the experimental setting and for generalizing results.

These properties provide new capabilities for CATE analysis, beyond the use of tabular surveys \citep{daoud_impact_2024,wager2018estimation,atheyStateAppliedEconometrics2017}. They enable the analyst and policymakers to see which units are responding better to the treatment in sample \citep{thejerzak2023image}, and also likely reveal critical information as to how the results should be generalized out-of-sample. This generalization is called transportability analysis \citep{pearl2022external}. Although there are several computer vision methods that use images for causal inference  \citep{sakamoto_planetary_2024,ratledge_using_2022,jerzak2023integrating,castroCausalityMattersMedical2020,pan_counterfactual_nodate}, little is known about the relative strengths and weaknesses of each approach when using sequences of satellite images to estimate CATE and subsequently to inform transportability in RCTs. 

This paper develops a methodological approach that addresses these two challenges. By \textit{methodological approach}, we mean a set of computer vision techniques for which their model complexity can be adapted to sample size, compute limitations, and other considerations. 

We develop our approach using both simulations and replicate two RCTs to show the strengths and weaknesses of these techniques. The purpose of the simulation is to show the extent to which different computer vision techniques are able to capture underlying signals in satellite image sequences while defining computational requirements. Although defining these requirements is challenging, we analyze the number of parameters each technique requires, something correlated with compute requirements (in terms of, e.g., GPU memory).

The first RCT is drawn from the Multifaceted Graduation Program in Peru \citep{banerjee2015}, with randomization of treatment occurring in 2011. In this application, we study household poverty as it responds to an educational intervention in the Cusco region of Peru. Our techniques decompose those responses by image sequence---with this sequence containing information on the developmental trajectory of each neighborhood. In the second RCT, we revisit a topic of growing interest in the climate change community---adaptation to extreme climactic events \citep{bolsen2014voters}. These applications contrast well both in terms of how satellite images matter for climate and poverty issues and in terms of sample size. While the Peru setting contains only 86 villages, the Georgia case has 3,362 geographic areas--a sample size difference that will be illuminating in terms of EO-based CATE decomposition.

The remainder of this paper is structured as follows. In Section \ref{s:Models}, we introduce our methodological framework for combining causal inference with computer vision techniques for satellite imagery. Section \ref{s:Sim} presents a simulation study to evaluate the performance of our proposed methods under controlled conditions where we have full knowledge of the process by which the image sequence information translates into a heterogeneity response. Section \ref{s:TwoRCTs} applies our approach to two real-world RCTs in Peru and Georgia, demonstrating its practical utility and providing insights into effect heterogeneity and transportability. In Section \ref{s:Conclusion}, we discuss implications of our findings, limitations of our approach, and directions for future research before concluding. Throughout, we emphasize both the potential and challenges of using earth observation data for causal inference in diverse experimental settings. We make all modeling approaches available in an open-source package available at \href{https://github.com/AIandGlobalDevelopmentLab/causalimages-software}{\url{GitHub.com/AIandGlobalDevelopmentLab/causalimages-software}}.

\section{Combining Causal Inference and Computer Vision for Satellite Image Sequences}\label{s:Models}

In the experimental context, image-sequence information may contain insight into how units will respond to the intervention. For example, in the earth observation context, pre-treatment image-sequence data contains information about developmental trajectories in the neighborhood context \citep{pettersson2023time,daoud_using_2023-1,yeh2020using}. Therefore, conditioning on pre-treatment image-sequence data can allow researchers to decompose effect heterogeneity on the basis of historical trajectories, in addition to static neighbor dynamics only. 

We denote the binary treatment variable as $W_i\in\{0,1\}$; potential outcomes are denoted by $Y_i(0)$ and $Y_i(1)$, representing outcomes if unit $i$ receives treatment or control intervention. We denote a single pre-treatment image array associated with unit $i$ as $\bM_i \in\mathbb{R}^{W\times H \times B}$, with image width/height axes $W,H$ and bands $B$. Examples of bands would be the red, green, and blue frequency ranges, together with frequency ranges in the UV and infrared domains. $\bM_i$ takes on a specific realization $\mathbf{m}$. Likewise, we denote a pre-treatment image sequence as $\bV_i \in\mathbb{R}^{T\times W\times H \times B}$ (with specific realization $\bv$) and $\bX_i$ denotes pre-treatment tabular covariates such as age and income (with specific realization $\mathbf{x}$). We are interested in understanding and comparing the dynamics of the following quantities, separately and jointly: 
\begin{align*}
\tau(\mathbf{x}) &= \E[Y_i(1) - Y_i(0) \; \mid \; \bX_i  = \bx] \;\; \textrm{\it (Tabular CATE)}
\\ \tau(\mathbf{m}) &= \E[Y_i(1) - Y_i(0) \; \mid \; \bM_i  = \mathbf{m}] \;\; \textrm{\it (Image CATE)} 
\\\tau(\bv) &= \E[Y_i(1) - Y_i(0) \; \mid \; \bV_i  = \bv] \;\; \textrm{\it (Video CATE)}
\\\tau(\bv, \bx) &= \E[Y_i(1) - Y_i(0) \; \mid \; \bV_i  = \bv, \bX_i  = \bx] \;\; \textrm{\it (Joint Tabular+Video CATE)}
\end{align*}

Use of tabular data for CATE estimation is well-studied. There is a wealth of literature studying heterogeneous treatment effects by age \citep{dahabreh2016using}, gender \citep{kent2008gender}, and ethnic subgroups \citep{adida2017overcoming}, to list just a few important moderating factors that have been subject to careful examination by scholars across economics, epidemiology, political science, sociology, and related fields.

While collecting tabular data for CATE estimation will remain important, several limitations are worth mentioning. First, tabular CATEs can be expensive to obtain and are often acquired via surveys that can cost upwards of \$13 to \$400 per respondent, independently of the cost of treatment \citep{ripley2010s}. The integration of experimental data with proprietary data such as the L2 voter files can also be expensive, with prices ranging from \$0.25 per person-entry to \$37,000 per US state \citep{ballotpedia_voter_files}. More conceptually, tabular CATEs are limited in the degree to which they can inform about transportability. If covariates were collected in the experiment only, then non-experimental sites lack key covariates upon which to base generalizability exercises, especially across national or subnational boundaries and where administrative data may differ. 

To address these limitations, new methodologies have emerged that use raw image data for CATE estimation \citep{jejoda2022_hetero}. In particular, image CATEs can be informative in breaking down effect heterogeneity by neighborhood or other contextual features of an area, such as the environmental characteristics, urbanization, arrangement of transportation network \citep{nagne2013transportation}, and agricultural patterns \citep{holmgrenSatelliteRemoteSensing1998a}. In this article, we introduce the estimation of video CATEs, which extends contextual heterogeneity decomposition into the spatio-temporal domain. With our approach, we can not only analyze fixed features of an experimental unit's context but also that context dynamically evolving.

We now turn to a discussion of how this evolution may be quantified in the context of RCTs.

\subsection{Choice of Image Sequence Model}\label{s:ModelsIntro}

Our purpose here lies in quantifying the influence of different image sequence choices on subsequent effect heterogeneity analysis. We, therefore, compare a set of five image sequence architectures, $f$, all of which take the image sequence and output a vector representation: 
\begin{align*}
f: \bV_i \in \mathbb{R}^{T\times W\times H\times B} \mapsto \mathbb{R}^D, 
\end{align*}
where $D$ denotes the dimensionality of the output representation (typically much smaller than the input dimensionality of $T\times W\times H\times B$). Here, $T$ represents the number of time slices, $W$ the image width, $H$ image height, and $B$ number of frequency bands. For all modeling strategies, we compare performance in the image sequence space with the image-only and tabular-only cases.

\begin{figure}[ht!] 
 \begin{center}  
\includegraphics[width=0.95\linewidth]{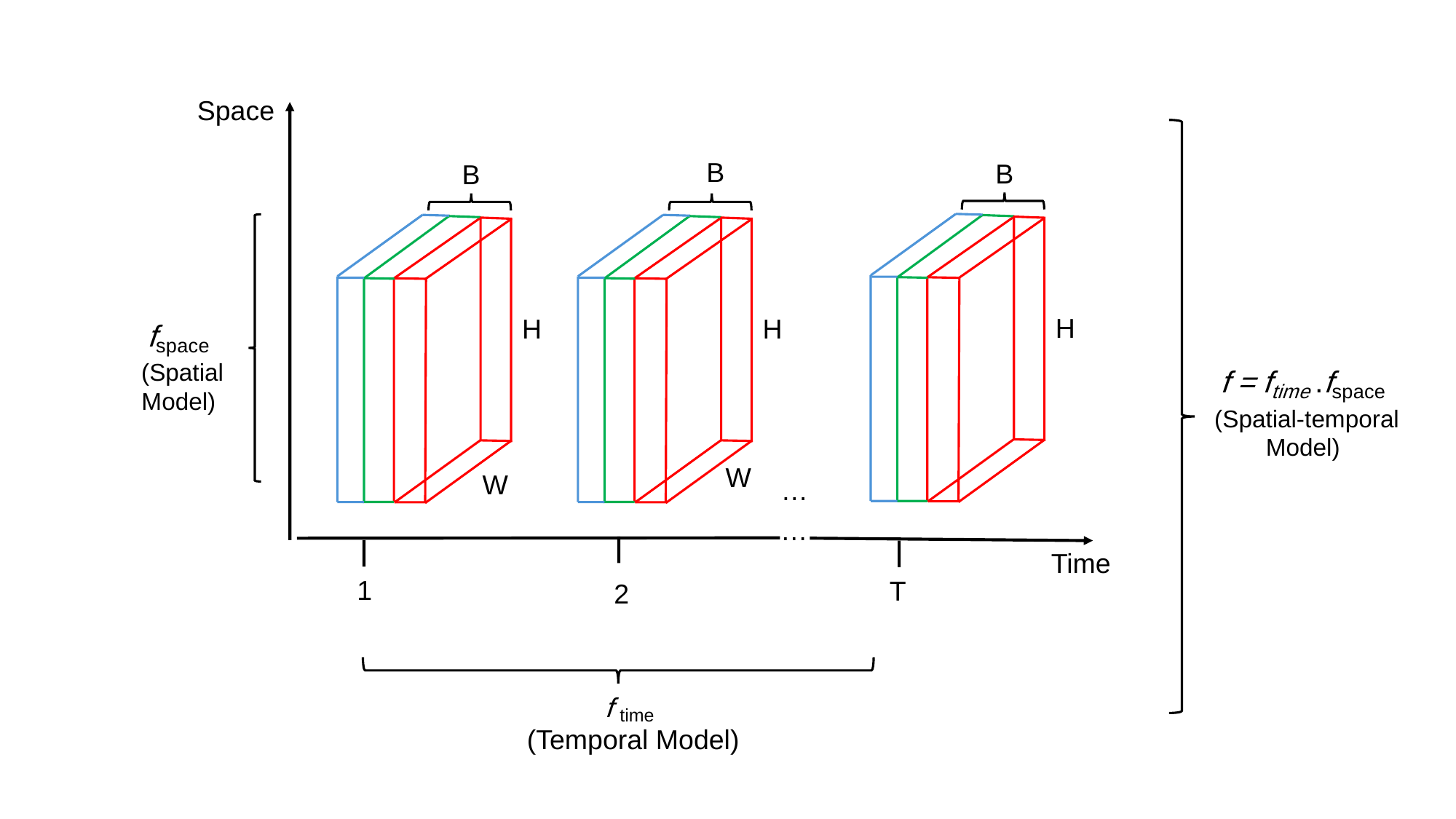}
\caption{
Visualization of the image sequence data, where modeling strategies must summarize across spatial and temporal information, either separately or jointly.
}
\label{fig:DataViz}
\end{center}
\end{figure}

In Figure \ref{fig:DataViz}, we see the structure of the input image sequence arrays. There is a spatial dimension for these images ($H$ and $W$), and a temporal dimension ($T$) representing how the image evolves over time. In principle, we could either (a) first summarize this image information spatially then temporally, (b) first temporally then spatially, or (c) jointly (i.e., spatio-temporally). For computational reasons, we will primarily consider approaches that first summarize across space (marginalizing across the $W\times H$ dimension) and then across time (marginalizing across the $T$ dimension). 

\begin{figure}[ht!] 
 \begin{center}  
\includegraphics[width=0.95\linewidth]{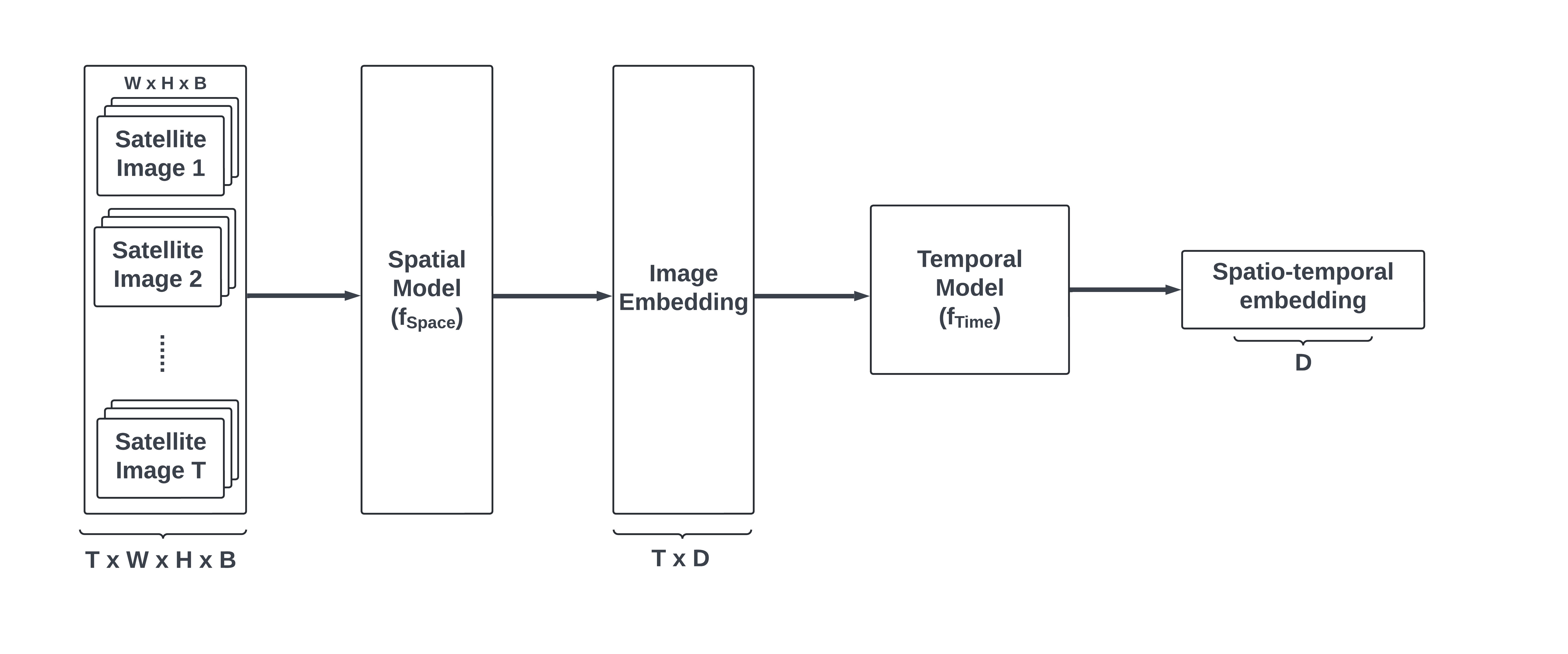}
\caption{
Visualization of the spatio-temporal embedding process for unit $i$ (excluding the Video Autoencoder). Green nodes denote raw data; blue nodes denote models; gray nodes denote embeddings. In the two experiments, $T=$ 5.
}
\label{fig:EmbeddingsFlow}
\end{center}
\end{figure}

This spatial-then-temporal approach is visualized in the middle and right panels of Figure \ref{fig:EmbeddingsFlow}, where one model usually summarizes the spatial characteristics of each image slice, and another model captures the temporal characteristics across the image sequence. Thus, image sequences can be efficiently modeled by a function, $f$, with two constituent parts: one spatial, $f_{Space}$, and another temporal, $f_{\textrm{Time}}$. Those two components then build up our EO-ML model $f$ by composition, $f=f_{\textrm{Time}}\circ f_{\textrm{Space}}$. However, there are many possible computer vision models that can populate the spatial $f_{\textrm{Space}}$ block and time series models for the temporal $f_{\textrm{Time}}$ blocks. 

Because there are a large number of models suitable for occupying the role of $f_{\textrm{Space}}$ and $f_{\textrm{Time}}$, we restrict our selection to three primary architectural classes that differ in parameter number and training data (no training vs. consumer image/video training data vs. EO training data).

\textit{Class 1} uses randomized projections to summarize the image sequence arrays into a vector representation through a randomized neural map \citep{rolf2021generalizable}; we compare CNN and Vision Vision Transformers for summarizing spatial representations. In both cases, a temporal Transformer summarizes across the temporal domain into the final output. Intuitively, in this approach, a neural network is initialized, but not trained, with the output containing a (randomized) compressed representation of the original input that can be informative when fed into a downstream model for CATE prediction. 

\textit{Class 2} relies on pre-trained image and image sequence models that have used consumer image and video data in training. We first consider a Vision Transformer trained on images of objects from the ImageNet corpus; in this case, a randomized Transformer map summarizes across the temporal dimension. We also consider a full Video Auto Encoder trained on a large corpus of video data from Something-Something-V2 [sic] corpus. 

Finally, \textit{Class 3} employs an EO foundation model \citep{ClayFoundationModel} to generate image representations (and a randomized neural map to summarize the temporal slices into a final vector representation). Whereas the first model class uses no neural training and the second uses training on non-EO data streams, the model approach is unique in that the representation function is trained using a large corpus of EO data from Landsat and Sentinel satellites, among other sources. Appendix V contains a more detailed description of the model classes just described. Table \ref{tab:ModelSummary} summarizes the model classes explored in this work. 

\begin{table}[H]
  \centering
\renewcommand{\arraystretch}{1} 
  \caption{Estimation models for image sequences in causal EO-ML pipeline. Input: Image sequences, $\mathbb{R}^{T\times W\times H\times B}$. Output: Vector representations of the image sequences, $\mathbb{R}^D$.
   ``R'' denotes randomzied projection; ``PT'' denotes pre-trained. ``$+$'' denotes the sequential application of one model and another (e.g., to first aggregate across spatial and then across temporal information). ``ViT'' denotes Vision Transformer;  ``T'' denotes temporal Transformer; ``CNN'' denotes Convolutional Neural Network, and ``Clay'' the Clay Foundation model. ``PCs'' denotes principal components. We apply these model classes both to image (``M'') and image sequence (``V'') data. 
  }
\label{tab:ModelSummary}
\begin{tabular}{lllll}
\hline \hline
{\bf Model Class} & {\bf Model Description} & {\bf Training Data} & {\bf Parameter \#} & {\bf $D$}\\
\hline
{\it Randomized} & R-ViT (+R-T) & None & 0.6M & 128\\
{\it projections} & R-CNN  (+R-T) & None & 9M & 384 \\
& & & & \\ 
{\it General image (seq.)} & PT-ViT (+R-T) & ImageNet & 20M & 768 \\
{\it foundation models} & PT-Video Autoencoder & Something-Something-V2 [sic] & 86M & 768 \\
& & & & \\ 
{\it EO foundation model} & PT-Clay (+R-T) & Landsat, Sentinel & 111M & 768 \\
\hline
\end{tabular}
\end{table}

In principle, an end-to-end model could be trained to model heterogeneity dynamics from the image sequence data, but RCTs often have fewer than 1,000 distinct household locations. In cluster randomized trials, even fewer locations may be available. This constraint limits the potential of end-to-end causal models for image sequence data streams in RCTs. For example, the Clay model was trained on 34 terabytes of EO data; only 7 gigabytes of satellite images are available around the larger of our two applications in the Georgia experiment context. Hence, there is a need for modeling strategies that can work well even in the smallest RCTs.\footnote{We note potential in the observational case where there is greater access to geo-located treatment and outcome data.}

To serve this need, all three model classes described here can be readily implemented even in small experiments with fewer than 100 distinct village clusters, as we later illustrate with the experimental analysis of Peru. For all three model classes, we compare results using the raw representations ($D$ usually 768) and their principal components ($D=10$). We reserve a more detailed analysis of the principal component results for Appendix II. Appendix IV contains a glossary covering some of the abbreviations used. We also compare results when applying the models just described to tabular-only and image slices only (instead of image sequences). The latter comparison will help quantify the relative importance of the historical information contained in satellite images over time. 

\subsection{Assessing Heterogeneity Given Data and Model}\label{s:RATEIntro}

To provide guidance on how to use EO data for RCT analysis, we discuss how each of our models performs in practice, focusing on capturing effect heterogeneity in image sequences (i.e., video). It will be difficult to evaluate this performance question in real RCT data due to how true CATE estimates are unobserved. Still, we can gain some traction using the Rank-Weighted Average Treatment Effect (RATE) \citep{yadlowsky2021evaluating}. This measure allows us to quantify and compare the performance of different CATE estimation methods in the absence of ground truth individual treatment effects.

For a video conditioning variable, $\bV_i$, the RATE is defined as
\begin{equation}
  \textrm{RATE} = \int_{0}^1 \alpha(q) \bigg(
  \underbrace{\E[Y_i(1)-Y_i(0) \mid F(S(\bV_i)) \geq 1 - q]}_{\textrm{ATE among top $q$-th percentile under rule $S$}}
  -
  \underbrace{\E[Y_i(1) - Y_i(0)]}_{\textrm{Baseline ATE}}  \bigg)\; \textrm{d} q, 
\end{equation}
Let us unpack this equation. First, the scoring function $S(\cdot)$ encodes a prioritization rule. In the image sequence setting, this rule is defined as a mapping from the video space to the real space $\bV_i \mapsto \mathbb{R}$, such that our sample of experimental units, $\{1,2,...,n\}$, is prioritized to receive the treatment in decreasing scoring order, $S(\bV_i)$. The order is chosen as:
\begin{equation}
i_1, \;i_2, \;...,\; i_n \textrm{ s.t. }
S(\bV_{i_1}) \geq S(\bV_{i_2}) \geq \cdots \geq S(\bV_{i_n}).
\end{equation}
This process implies that unit $i$ with the largest value of $S(\bV_i)$ will be treated first (i.e., as $i_1$) and the unit with the lowest value will be treated last (i.e., as $i_n$). The priority scoring function $S(\cdot)$ is specified by the investigator; it is generated from a data sample independent from that used to calculate the treatment effect estimates to ensure statistical validity. We define $S(\cdot)$ here so that the RATE captures the difference in average treatment effect if subsetting the sample to the likely high responders and computing the average effect among that subgroup, averaging across the range of what constitutes a predicted high responder.

The RATE calculation proceeds using $F(\cdot)$ as the cumulative distribution function (CDF) of $S(\bV_i)$. Thus, when $q$ is small, we effectively subset the data to top percentiles (i.e., $1-q$ fraction of the population).\footnote{Here, we assume the output of $S(\bV_i)$ to be continuous and that there are no ties among units in their scoring values.} The difference between $Y_i(1)$ and $Y_i(0)$ for the prioritization score quantiles and the baseline ATE establishes the targeting operator characteristic (TOC), the lift from prioritizing treatment allocation based on expected treatment benefit as predicted from $\bV_i$. The RATE can thus be seen as where the overall lift, averaged across the range of prioritization group sizes. Large positive values of the RATE measure indicate the strong presence of detected CATEs under the prioritization rule; values near zero indicate either no detected CATEs under the data and model or an ineffectual prioritization rule. 

We compare two popular weighting mechanisms in the RATE calculation, Area Under the Treatment Effect Curve (AUTOC) and QINI, which differ somewhat in their approach and interpretation. 

AUTOC assesses the cumulative gain in treatment effect lift over the ATE as we treat more individuals, sorted by their predicted treatment effect. In this case, identity weighting is used when integrating over $q$ in the difference between ATE among the top $q$-th percentile and the baseline ATE: 
\begin{equation}
\alpha_{\textrm{AUTOC}}(q) = 1
\end{equation}
QINI weighing, on the other hand, measures the area between the QINI curve (which plots the cumulative treatment effect against the proportion of the population treated) and the 45-degree line. QINI up-weights the lift from the prioritization rule, with weights proportional to the quantile under consideration: 
\begin{equation}
\alpha_{\textrm{QINI}}(q) = q
\end{equation}
By weighting the integrand by $q$, we effectively adjust the weighting to cancel out how highly prioritized units occur more frequently in the left-hand side conditioning statement than others; QINI weighting thus ensures an equal contribution of each unit to the overall RATE calculation. Because of this fact, QINI emphasizes the improvement over random assignment and shows greater power to detect effects when heterogeneity is diffusely distributed through the population. By contrast, AUTOC focuses on the overall effectiveness of targeting and shows greater power when a small group experiences heterogeneous effects. 

For our purposes, we present results using the \RATEWT{} choice of $\alpha(q)$, but report results for \RATEWTALT{} weighing in Appendix II. We make this choice under the conservative assumption that EO-based heterogeneity is restricted to a small subgroup of the experimental population. We average RATE ratios over 5 cross-folds in the sample-splitting for $S(\cdot)$, reducing sample size but ensuring statistical validity. 
 
\section{Simulation}\label{s:Sim}

To analyze how different modeling strategies perform at capturing heterogeneity dynamics using RATE metrics, we first turn to simulation. In this simulation, we will vary the noise level in the potential outcomes, as well as the image or video modeling strategy used, to provide insight into the behavior of the various EO-ML models for analyzing experimental data. 

\subsection{Simulation Design}

Images consist of high-dimensional arrays, $\bM_i$, and image sequences, $\bV_i$, repeat those dimensions over $T$ time steps. As a result, there are many ways effect heterogeneity may manifest in image sequences, $\bV_i$. For example, treatment of improving road infrastructure to reduce poverty may be more or less impacting mountain regions due to remoteness. Thus, mountains will moderate the treatment effect, but mountains may have many different image manifestations making it hard to encode those in a simulation. 

In pursuit of interpretability, we simplify the problem: the simulation contains two possible treatment effects, $1$ and $-1$, with two time periods in the image sequences. From time 1 to 2, an image rotation may occur that determines CATE for each unit.  

More precisely, to generate heterogeneity dynamics in image sequences, we adopt the following causal model for the potential outcomes, $Y_i(w)$, with two cases depending on whether the image is not ($S_i=0$) or is ($S_i=1$) rotated: 
\begin{equation}\label{eq:Sim}
  Y_i(w) =
    \begin{cases}
      -w  + \epsilon_{i} & \text{in the absence of image rotation, } S_i=0\\
      w   + \epsilon_{i} & \text{in the presence of image rotation, } S_i=1\\
    \end{cases}       
\end{equation}
where $Y_i(w)$ denotes the potential outcome under treatment arm, $w\in\{0,1\}$ and $\epsilon_i$ a mean 0 Normal with variance $\sigma^2$. We vary $\sigma^2\in \{0.01, 0.1, 1\}$. $S_i$, the indicator for whether the image sequence associated with each observation, $\mathbf{V}_i$, rotates from period 1 to period 2, is selected randomly with probability 0.5. In this way, $\tau(\bv)=-1$ when $S_i = 0$ and $\tau(\bv)=1$ when $S_i = 1$, which follows from Equation \ref{eq:Sim}.

\begin{figure}[ht!] 
\begin{center}  
\includegraphics[width=0.25\linewidth]{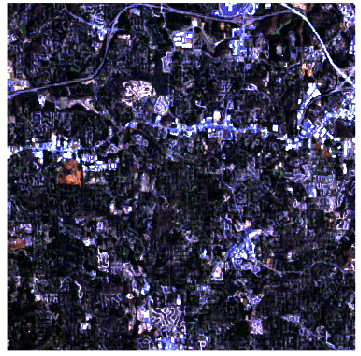}
\includegraphics[width=0.25\linewidth]{heterosim_n.png}
\caption{
  A responder for whom $S_i=0$ and $\tau(\bv) = -1$. Left and right images represent temporal periods 1 and 2.
}
\label{fig:SimNegResponder}
\end{center}
\end{figure}

\begin{figure}[ht!] 
 \begin{center}      \includegraphics[width=0.25\linewidth]{heterosim_n.png}
\includegraphics[width=0.25\linewidth]{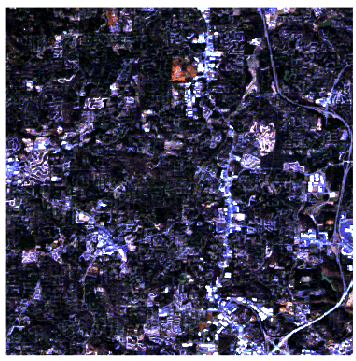}
   \caption{A responder for whom $S_i=0$ and $\tau(\bv) = 1$. Left and right images represent temporal periods 1 and 2. }
\label{fig:SimPosResponder}
\end{center}
\end{figure}

Figures \ref{fig:SimNegResponder} and \ref{fig:SimPosResponder} visualize image sequence examples that are associated with negative and positive responders. We draw images for this simulation study from Landsat satellite images drawn from the Georgia experimental sample analyzed in greater detail later. 

After generating representations from the various image sequence methods introduced in Section \ref{s:Models}, we then dispatch the resulting image sequence representations into a causal forest heterogeneity model as a flexible and widely used baseline \citep{athey2019generalized}. While we focus on this one class of heterogeneity model as our primary purpose here lies in exploring the role of different image sequence modeling strategies for effect heterogeneity, we compare results of the lasso-based R learner in Appendix I \citep{nie2021quasi}, which estimates treatment effects by first modeling outcomes and residuals before applying a learner to specifically model the heterogeneity process using covariates. 

We will first use an interpretable metric for capturing how well the different modeling strategies recover the true effect heterogeneity induced by the image sequences. In particular, our first metric of interest will be the bivariate correlation between estimated and true image sequence CATEs:
\begin{align*}
\textsc{CATE Quality Measure 1: } \textrm{Cor}\left( 
\widehat{\tau}(\bV_i),\;\tau(\bV_i)\right).
\end{align*}
When this correlation term is 1, the estimated CATEs match the true values in directionality; when 0, there is no systematic pattern relating the estimates to the truth. As a result, values close to 1 indicate success at capturing the video heterogeneity. Because this measure uses the true $\tau(\bV_i)$ values in its calculation, it can be only computed in simulation. 

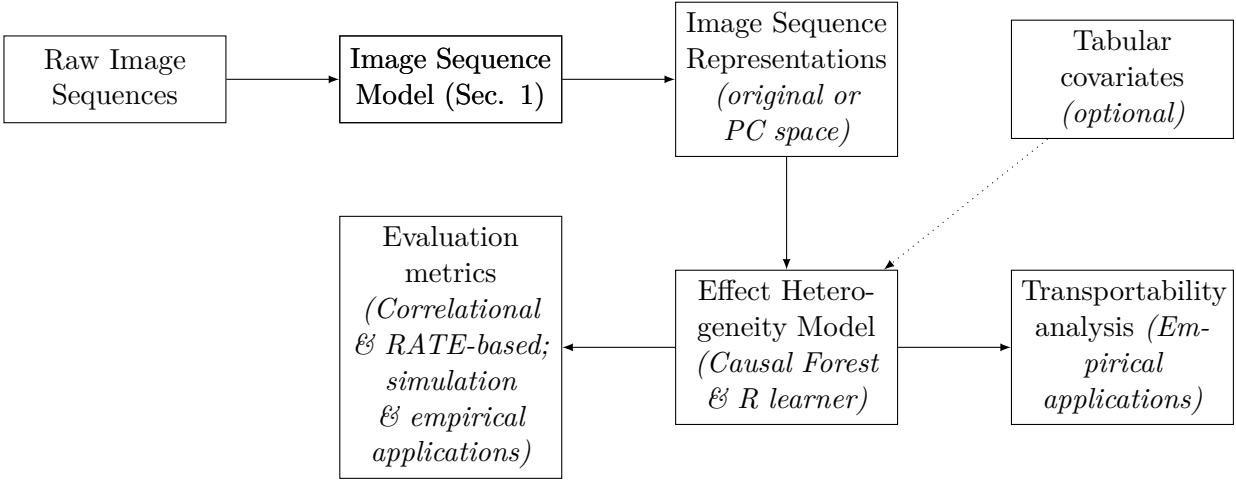
\begin{figure}[H]
\centering
\begin{tikzpicture}[node distance=1.5cm, auto,
    block/.style={rectangle, draw, text width=7em, text centered, minimum height=3em},
    line/.style={draw, -latex}]

    \node [block] (input) {Raw Image Sequences};
    \node [block, right=of input] (methods) {Image Sequence Model (Sec. \ref{s:Models})};
    \node [block, right=of input] (methods) {Image Sequence Model (Sec. \ref{s:Models})};
    \node [block, right=of methods] (representations) {Image Sequence Representations {\it (original or PC space)}};
    \node [block, below=of representations] (causalforest) {Effect Heterogeneity Model {\it (Causal Forest \& R learner)}};
    \node [block, right=of representations] (tabular) {Tabular covariates {\it (optional)}};
    \node [block, left=of causalforest] (metrics) {Evaluation metrics {\it(Correlational \& RATE-based; simulation \& empirical applications)}};
    \node [block, right=of causalforest] (transport) {Transportability analysis {\it(Empirical applications)}};
    
    \path [line] (input) -- (methods);
    \path [line] (methods) -- (representations);
    \path [line, dotted] (tabular) -- (causalforest);
    \path [line] (representations) -- (causalforest);
    \path [line] (causalforest) -- (metrics);
   \path [line] (causalforest) -- (transport);
\end{tikzpicture}
\caption{Pipeline for image sequence-based effect heterogeneity analysis.}
\label{fig:pipeline}
\end{figure}

As noted above, we are also interested in analyzing performance in a measure that can be computed in real experiments where no ground truth CATE information is available. Thus, we also analyze the following quantity, discussed in Section \ref{s:RATEIntro}: 
\begin{align*}
  \textsc{CATE Quality Measure 2 (``RATE Ratio''): } = \frac{\widehat{\textrm{RATE}}}{
\widehat{\textrm{s.e.}}(\widehat{\textrm{RATE}})}. 
\end{align*}
When the RATE ratio is greater than 1.96, there is a statistically significant heterogeneity signal given the data and model. When the ratio is near 0, there is little evidence of heterogeneity signal in the data and model setting. The ratio is comparable across experiments, as it is a scale-free measure resulting from dividing the point estimate by the standard error.  By analyzing CATE performance in this case, we will gain insight into the ability of these measures in the context where investigators seek to explore heterogeneity using image sequence information. Figure \ref{fig:pipeline} illustrates our full analysis pipeline.

\subsection{Simulation Results}

\begin{figure}[ht!] 
 \begin{center}  
   \includegraphics[width=0.45\linewidth]{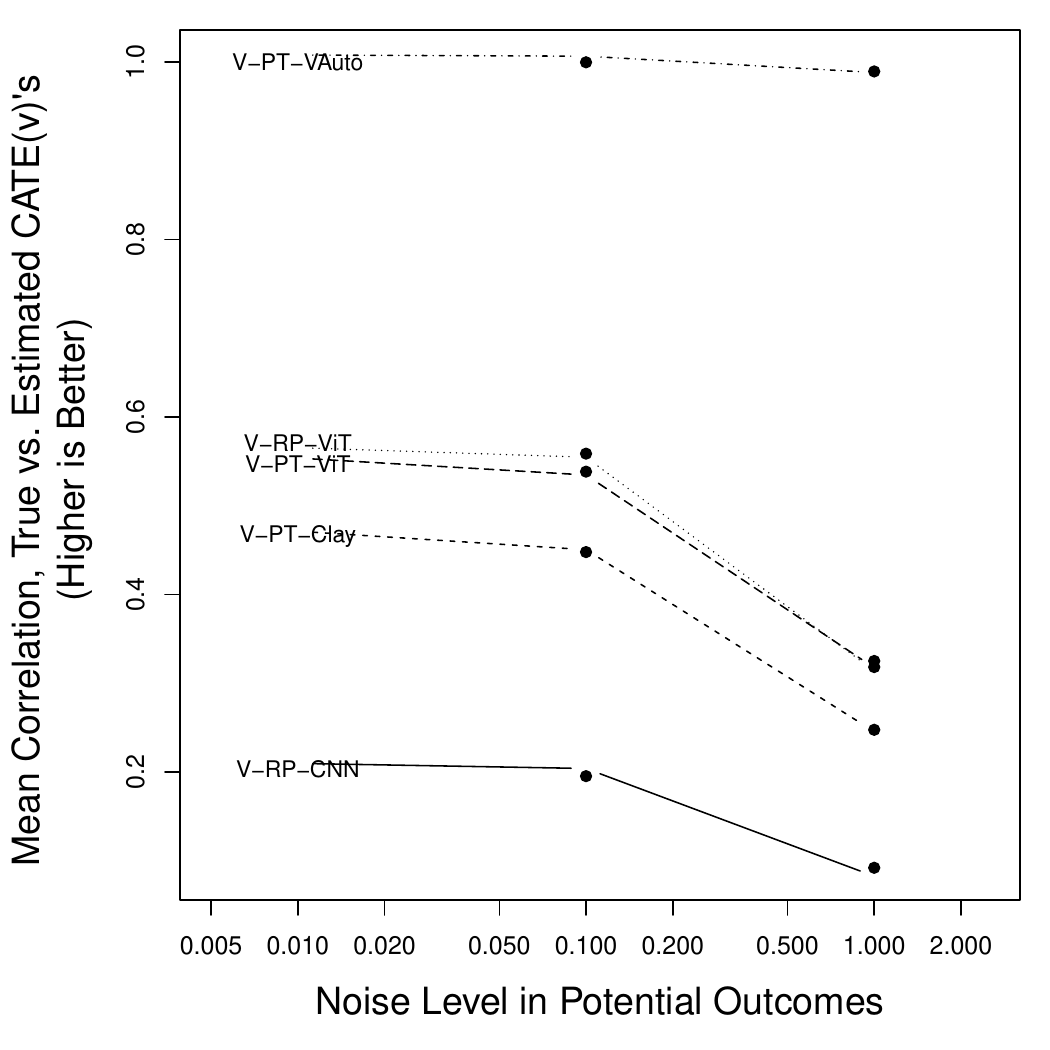}
   \includegraphics[width=0.45\linewidth]{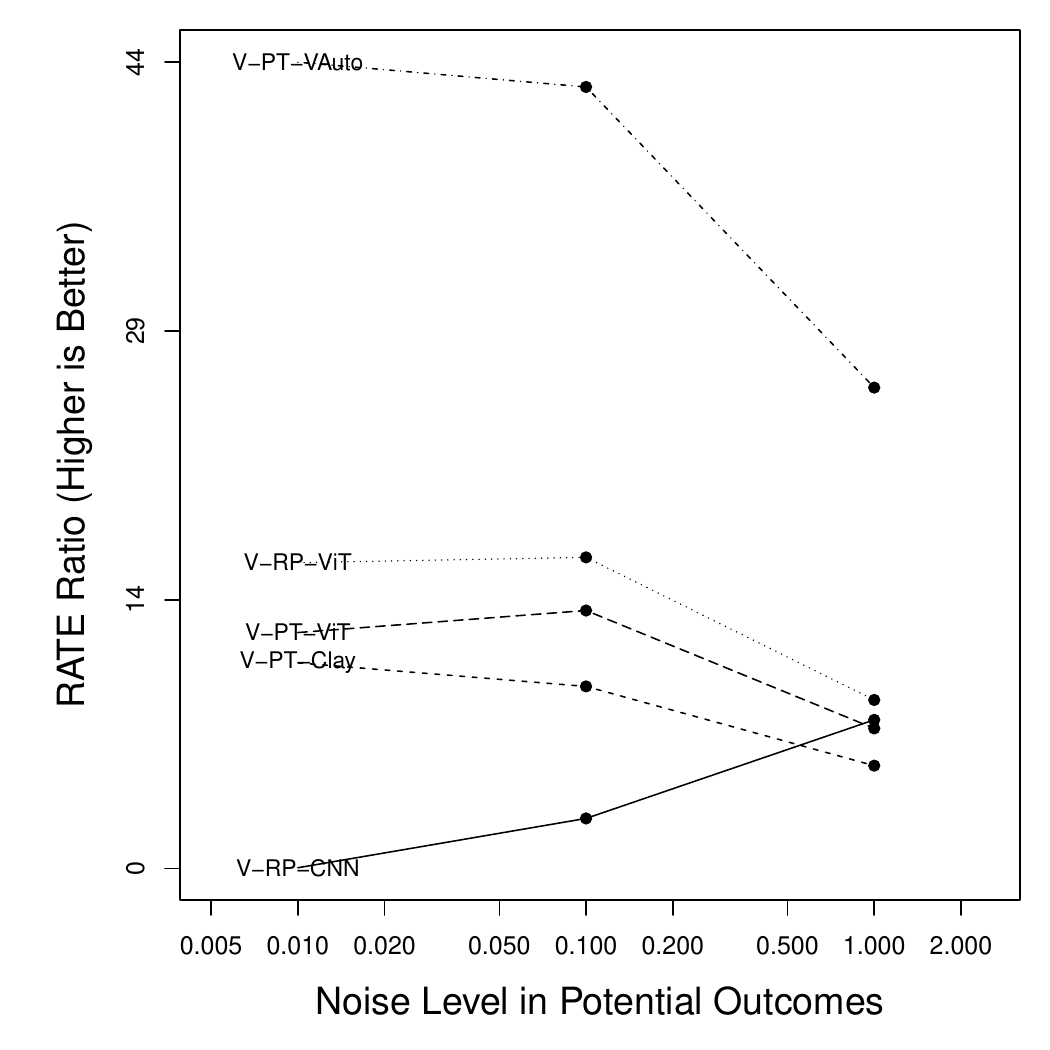}
   \caption{
     Simulation results.
     {\sc Left.} Correlation between true and estimated image sequence CATEs. 
     {\sc Right.} RATEs using AUTOC weighting. 
    In both panels, ``V'' denotes video, ``RP'' denotes randomized projection, ``CNN'' denotes Convolutional Neural Network; ``ViT'' denotes Vision Transformer. ``Vid'' denotes Video Masked Autoencoder. ``Clay'' denotes the Clay Foundation model.
   }
\label{fig:SimMain}
\end{center}
\end{figure}

The left panel of Figure \ref{fig:SimMain} presents the main simulation results using the correlation measure. We find that all methods show the ability to detect non-trivial CATEs in this case, with all correlations with the true CATEs falling above 0.15. As expected, increased noise in the potential outcomes reduces the correlation of estimated with true image sequence CATEs. The end-to-end video model shows best performance in this context, perhaps because of its focus on extracting features related to kinetics and motion, with the randomized Vision Transformer and pre-trained models showing similar and slightly lower performance. The Clay EO foundation model shows reasonable performance; the randomized CNN shows the weakest performance in this case, possibly because the CNN model extracts features related to texture, something not relevant in generating the CATE signal here. 

The right panel of Figure \ref{fig:SimMain} presents the main simulation results using the RATE measure, which is more applicable in practice given the models' ability to estimate RATEs in real RCTs. Interestingly, we find that the RATE ratio is not always statistically powerful in detecting heterogeneity even when there is a non-trivial correlation between estimated and true CATEs. This fact suggests that, even when the RATE measure fails to detect heterogeneity given data and model at a statistically significant level, the data and model combination may still be capturing real heterogeneity information. Nevertheless, when the true heterogeneity is well-modeled, as in the Video Masked Autoencoder approach, the RATE ratio is large ($>5$), indicating a pronounced heterogeneity signal. 

In Appendix I, we find reduced relative performance for the heterogeneity modeling approach using principal component reductions of the raw image sequence representations (Figure \ref{fig:SimMainPC}). We also find reduced performance of the lasso-based R learner methodology relative to the causal forest in this case (see Figures \ref{fig:SimR} and \ref{fig:SimRPC} for results from the full image sequence and principal component representations with the lasso-based R learner). 

Overall, these simulation results show that, under controlled conditions, it is possible to detect image-sequence-based heterogeneity even with a relatively small sample size (here, $n=1000$). We caution against reading too much into the relative ordering of the results here, as in EO settings, large rotations of the image from time period to time period are not possible. It is plausible, even likely, that in the EO-RCT context, different kinds of information other than kinetics are more relevant for treatment response processes. 

\clearpage 
\section{Empirical Analysis: Earth Observation Analysis of Heterogeneity in Two RCTs}\label{s:TwoRCTs}

To illustrate the strengths and limitations of EO-based experimental analysis, we consider two RCTs performed in disparate social and economic experiments---one in the United States and the other in Peru. The experiment in the US explores the effect of a public goods intervention in the context of climate change in the state of Georgia \citep{bolsen2014voters}. The experiment in Peru looks at the effect of a multifaceted graduation program on improvements in the well-being of people. Whereas the Georgia experiment was performed in an urban and suburban area with experimental units distributed evenly throughout the experimental context, the Peru experiment was performed in small villages, with units clustered tightly in space. Thus, by comparing results from these two disparate cases, we will be able to gain insight into the settings where image and image sequence data from satellites provide a stronger signal into underlying causal processes.

\subsection{A Climate Change Experiment in the United States}\label{s:GeorgiaIntro}

This experiment tested whether pro-water conservation messages affected monthly household water consumption in Cobb County, Georgia. It was performed between June and September 2007. The treatment variable is random assignment of households to various types of pro-water conservation messages. There were a total of 106,872 households as part of the experiment, where 35,093 households were assigned to the treatment group, and the remaining 71,779 households were assigned to control. Households in the treatment group were administered three versions of the water conservation message, with increasing magnitude of social encouragement: (1) an information-only tip sheet on water conservation, (2) the tip sheet plus a pro-social appeal to conserve water, and (3) the tip sheet, the pro-social appeal, and a social comparison of household water usage, with the assumption that each subsequent treatment would lead to progressively greater reductions in water consumption.

\begin{figure}[H] 
 \begin{center}  
\includegraphics[width=0.45\linewidth]{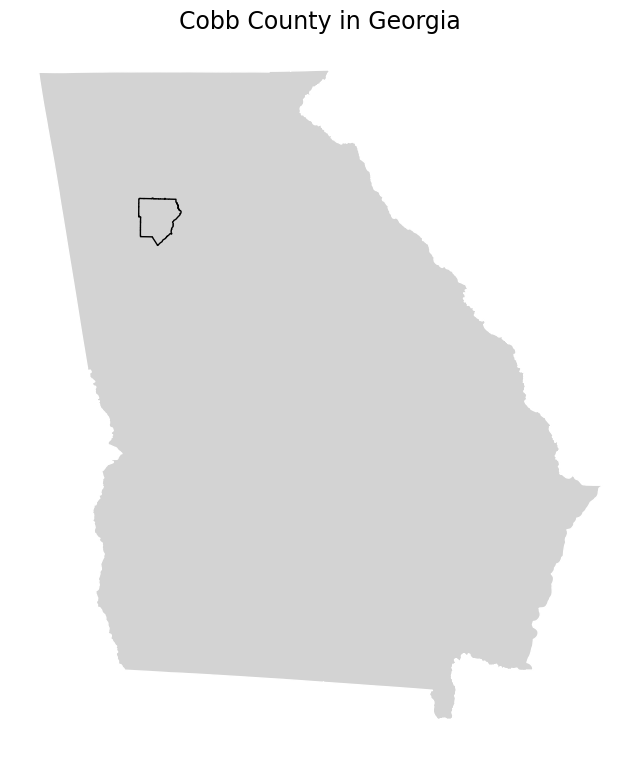}
\includegraphics[width=0.45\linewidth]{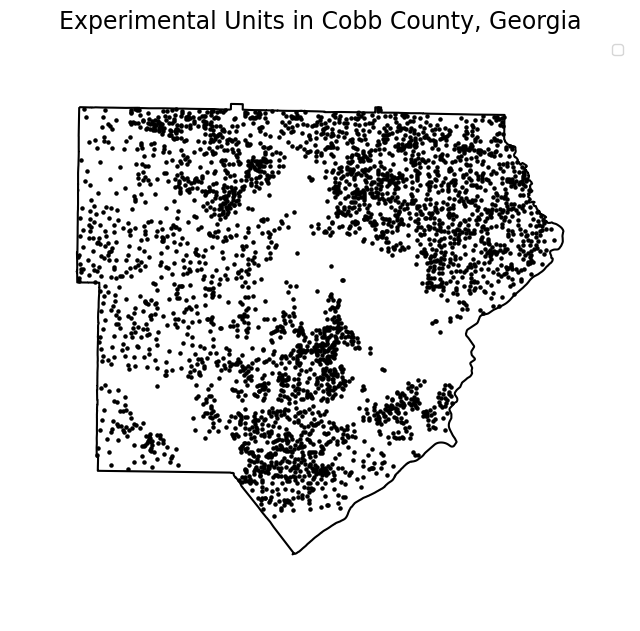}
\includegraphics[width=0.45\linewidth]{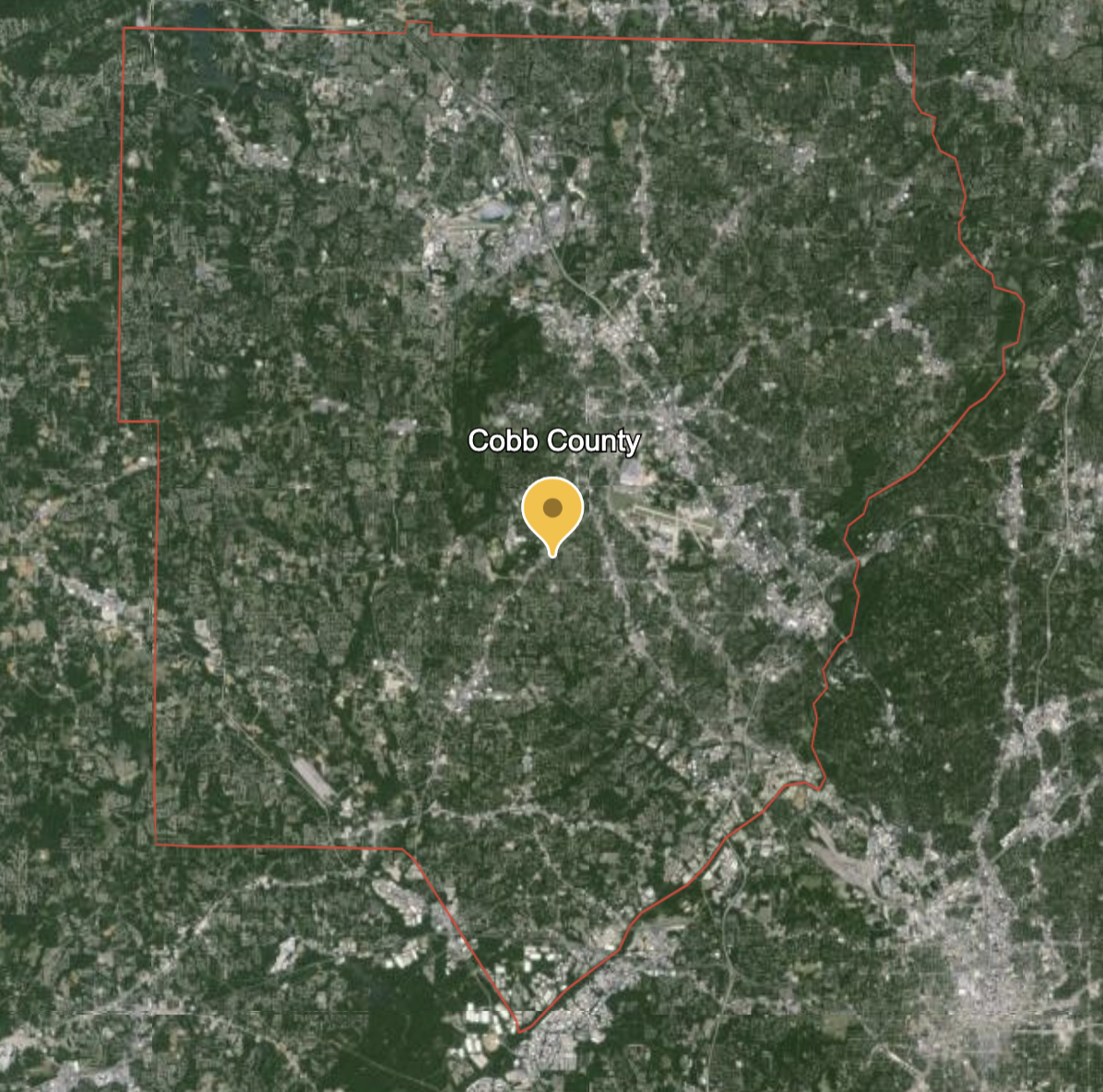}
\caption{
{\sc Left.} Georgia, with Cobb County indicated.  
{\sc Right.} Experimental units in Cobb County. 
{\sc Bottom.} Satellite image representation of Cobb County.
}
\label{fig:GeorgiaDescriptives}
\end{center}
\end{figure}

The outcome variable is a change in household water consumption (post-treatment to pre-treatment). Information about the locations of households is obtained from the household's census block. There are a total of 3,362 census blocks corresponding to 106,872 households in the data. Pre-treatment image data are taken from Landsat 5. This data includes a sequence of five median composite images with RGB bands taken every 2 years over a period of ten years before the experiment started for each household. For instance, the first image contains median pixel values taken between January 1, 1997 and December 1, 1998. The second image similarly has median pixel values between January 1, 1999 and December 1, 2000. In this way, we have a sequence of 5 images for each household. Since the location information and the satellite images correspond to census blocks, there is an overlap of images for households.

We also study individual-level covariate information (log age), as well as Census block-level information on the percentage of the given area that is urban, as well as the population proportion Black, Hispanic, with bachelor's degree, and living under the poverty line.

\subsection{Multifaceted Graduation Program in Peru}\label{s:PeruIntro}

The RCT in Peru was conducted around 2011 as part of a larger study done across multiple countries. The goal was to study the effect of a multifaceted graduation program on generating lasting improvements in economic well-being. Here, we only look at the data from Peru; we take our sample to be the observations for which outcome and location information are available (causing loss of 898 observations, mostly in the control group). 

The treatment variable is the random assignment of households to a combination of six interventions, including productive asset transfer, consumption support, and health education. Treatment is assigned using a clustered randomization scheme. A total of 86 villages in Peru were randomly selected to be either treatment or control villages, and then treatment households were randomly selected within the set of eligible households in treatment villages. We take household inclusion into a conditional cash transfer program as our binary treatment (the condition is that heads of households obtain identity cards for their children, take children to health checkups, and send children to school). 

Here, the outcome variable of interest is difference between a household asset index between endline and baseline. We obtain satellite images of households using the locations of individual households and Landsat 5, again using the pixel medians across 2-year time slices (excluding clouded pixels) starting from the year 2001.\footnote{When we take the median or mean over time dimension, we reduce variability over time, and thus potentially CATE detection.} For instance, the first image contains median pixel values between January 1, 2001 and December 1, 2002, with the second image containing median pixel values between the same period between 2003 and 2004. We also study individual-level covariate information about whether the head of household has formal education, as well as village-level information on village population (log scale) and on the Progress Out of Poverty Index (PPI).

\begin{figure}[H] 
 \begin{center}  
\includegraphics[width=0.37\linewidth]{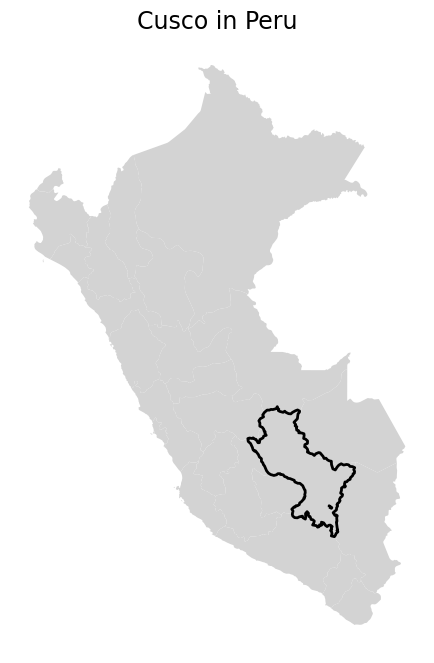}
\includegraphics[width=0.37\linewidth]{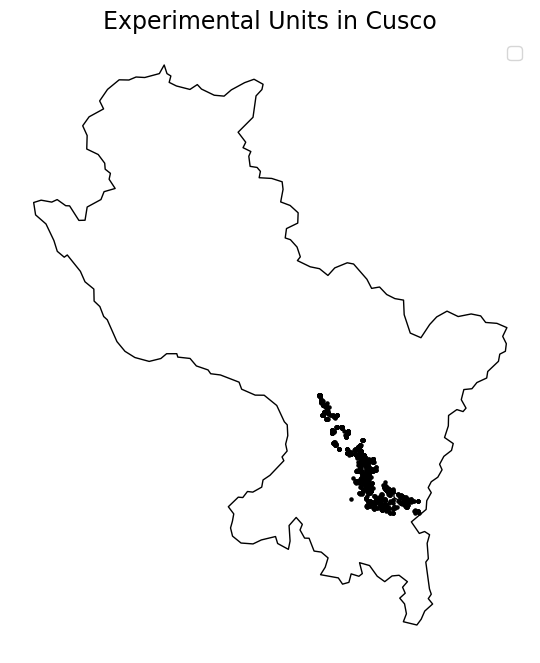}
\\ \vspace{0.75cm}
\includegraphics[width=0.37\linewidth]{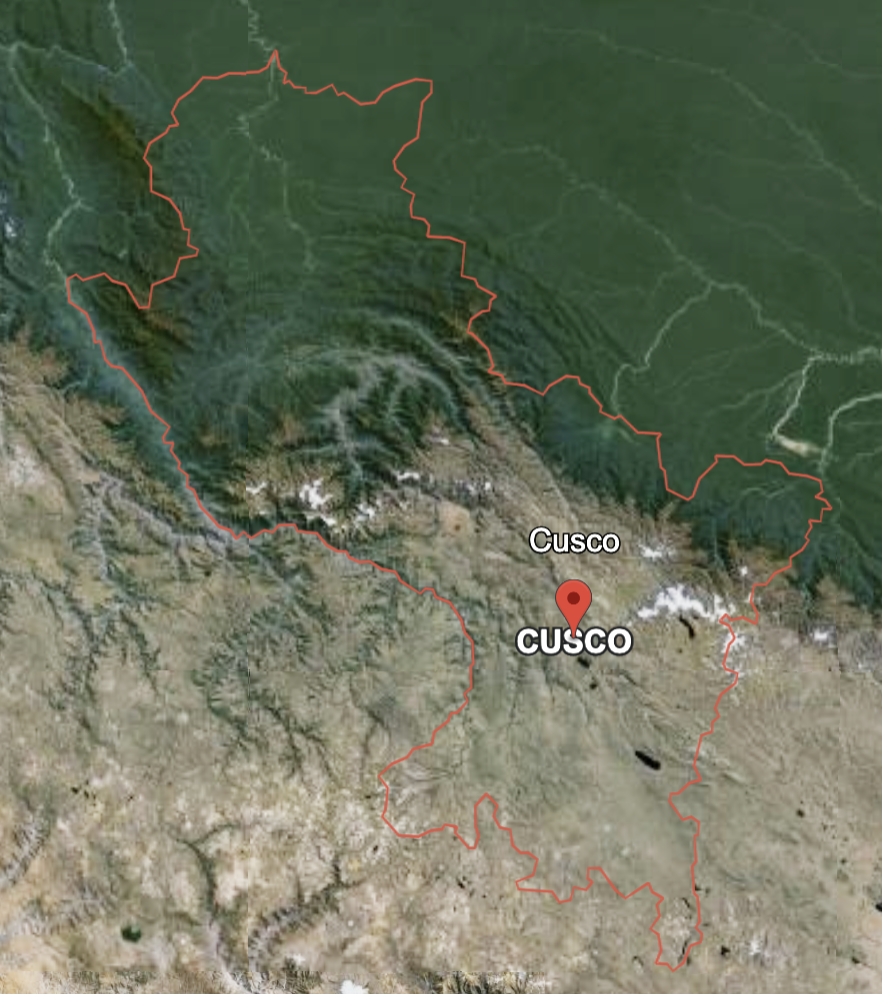}
\caption{
{\sc Left.} Peru, with Cusco indicated.  
{\sc Right.} Experimental units in Cusco. 
{\sc Bottom.} Satellite image representation of Cusco.
}
\label{fig:PeruDescriptives}
\end{center}
\end{figure}


In the next subsections, we first analyze the heterogeneity results from the experimental areas before comparing the image sequence results with those from land cover maps. We conclude this section with a transportability study, where we take the results from the experimental context and analyze, under assumptions, the spatial distribution of treatment effects derived using the pre-treatment satellite image sequences. 


\subsection{Effect Heterogeneity Analysis}

We now turn to our main heterogeneity analysis using the Georgia and Peru RCTs. 

We first examine the relationship between CATE estimates derived from tabular data, $\widehat{\tau}(\bX_i)$, and those from image and image sequence data ($\widehat{\tau}(\bM_i)$ and $\widehat{\tau}(\bV_i)$, respectively). Specifically, we calculate the Pearson correlation coefficient between these two sets of estimates. While correlation is a linear measure and may not capture all aspects of potentially non-linear relationships between CATEs, it provides a useful starting point for comparison.

The interpretation of these correlations requires unpacking. In some scenarios, researchers might expect or seek a high correlation. For instance, if treatment effect heterogeneity in an economic experiment is primarily driven by a few key factors (e.g., income, education) that are well-captured by both tabular and image data, we would expect similar CATE estimates across data sources. Conversely, a lower correlation could indicate that image sequence data is capturing novel sources of heterogeneity not present in tabular data. For example, satellite imagery might reveal spatial patterns of development or environmental factors influencing treatment effects that are not captured in traditional measured covariates such as age and pre-treatment income. 

\begin{figure}[H] 
 \begin{center}     \includegraphics[width=0.45\linewidth]{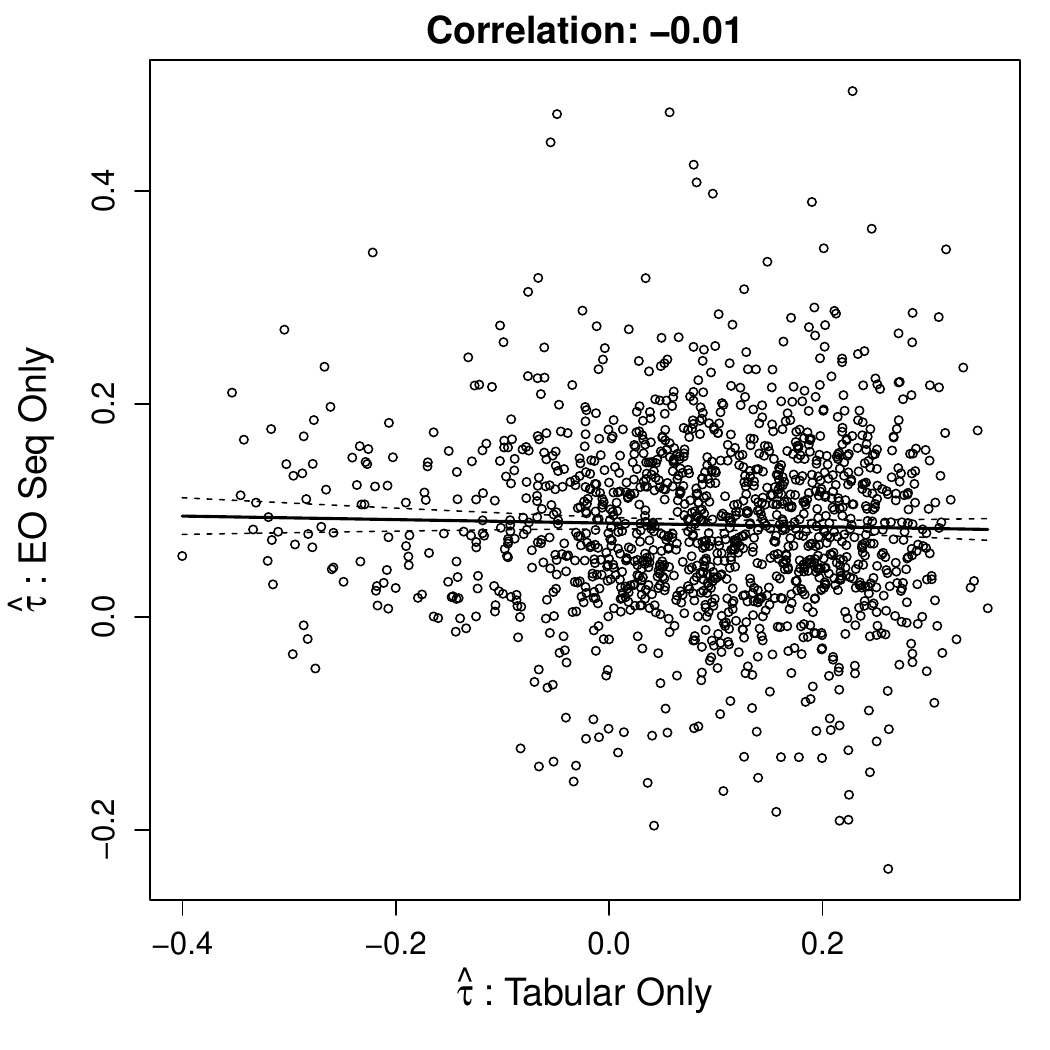}
   \includegraphics[width=0.45\linewidth]{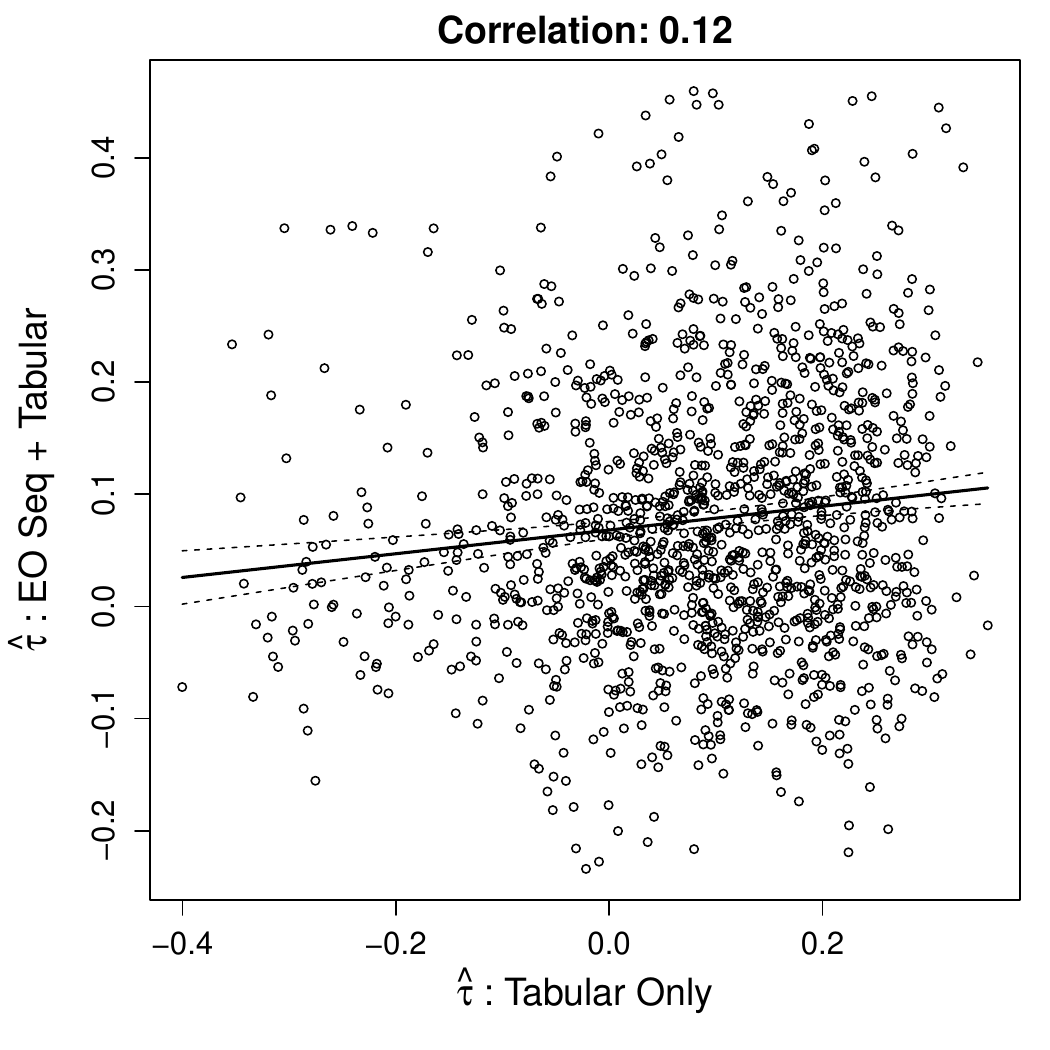}
   \includegraphics[width=0.45\linewidth]{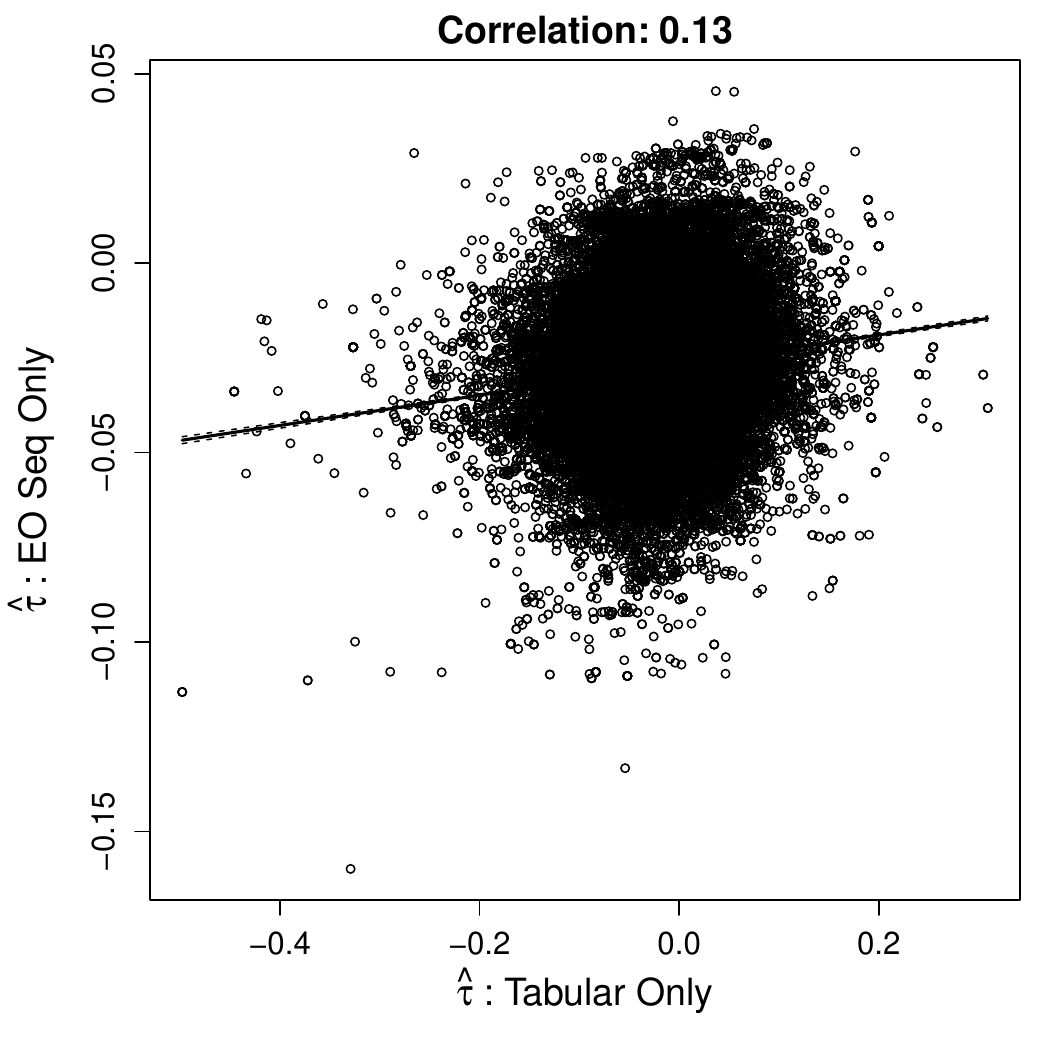}
   \includegraphics[width=0.45\linewidth]{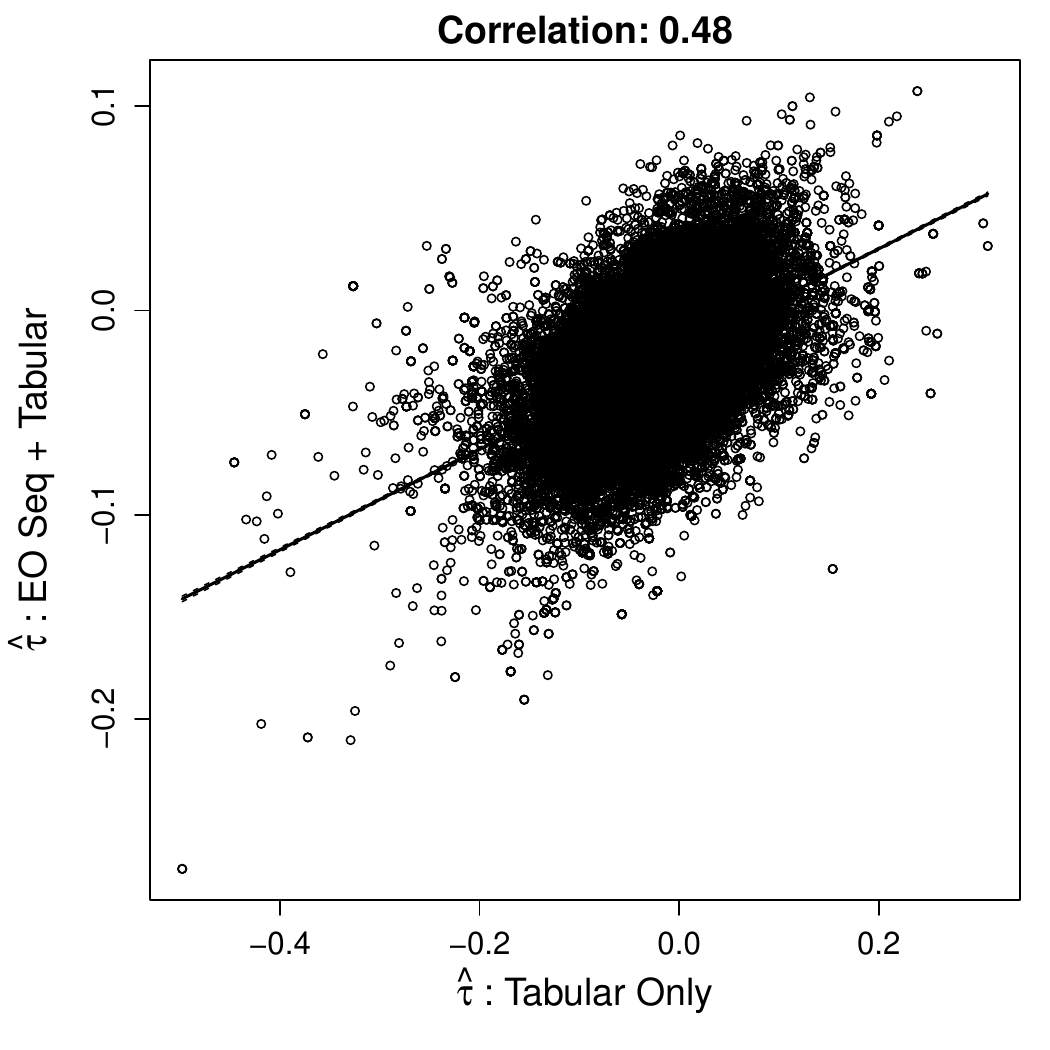}
   \caption{
{\sc Top.} Peru experiment.  {\sc Bottom.} Georgia experiment. 
  {\sc Left.} Correlation between CATEs estimated with tabular and with image sequence data using an EO foundation model. 
  {\sc Right.} Correlation between CATEs estimated with tabular and with image sequence data with tabular pre-treatment covariates added to the EO foundation model representations in the conditioning set. 
}
\label{fig:CorTabEO}
\end{center}
\end{figure}

With these qualifications in mind, we see in Figure \ref{fig:CorTabEO} the correlation between tabular CATEs and image sequences, both with (right) and without (left) tabular information appended to the image sequence representations. By tabular information, we refer to the individual- or group-level covariate features measured by experimenters such as age and Census block population (in the Georgia RCT) and household head education (in the Peru RCT) (see Section \ref{s:GeorgiaIntro} and \ref{s:PeruIntro} for more information). In the EO-only case, we note a higher correlation in the Georgia case, perhaps due to the fact that there are more unique sites in the Georgia experiment with which to expose heterogeneity dynamics. As expected, with tabular covariates appended to the image representations, the correlation with tabular-only CATEs increases. 

With these correlations in hand, we next proceed to an analysis of CATE quality using RATE analysis.

\begin{figure}[ht!] 
 \begin{center}  
\includegraphics[width=0.55\linewidth]{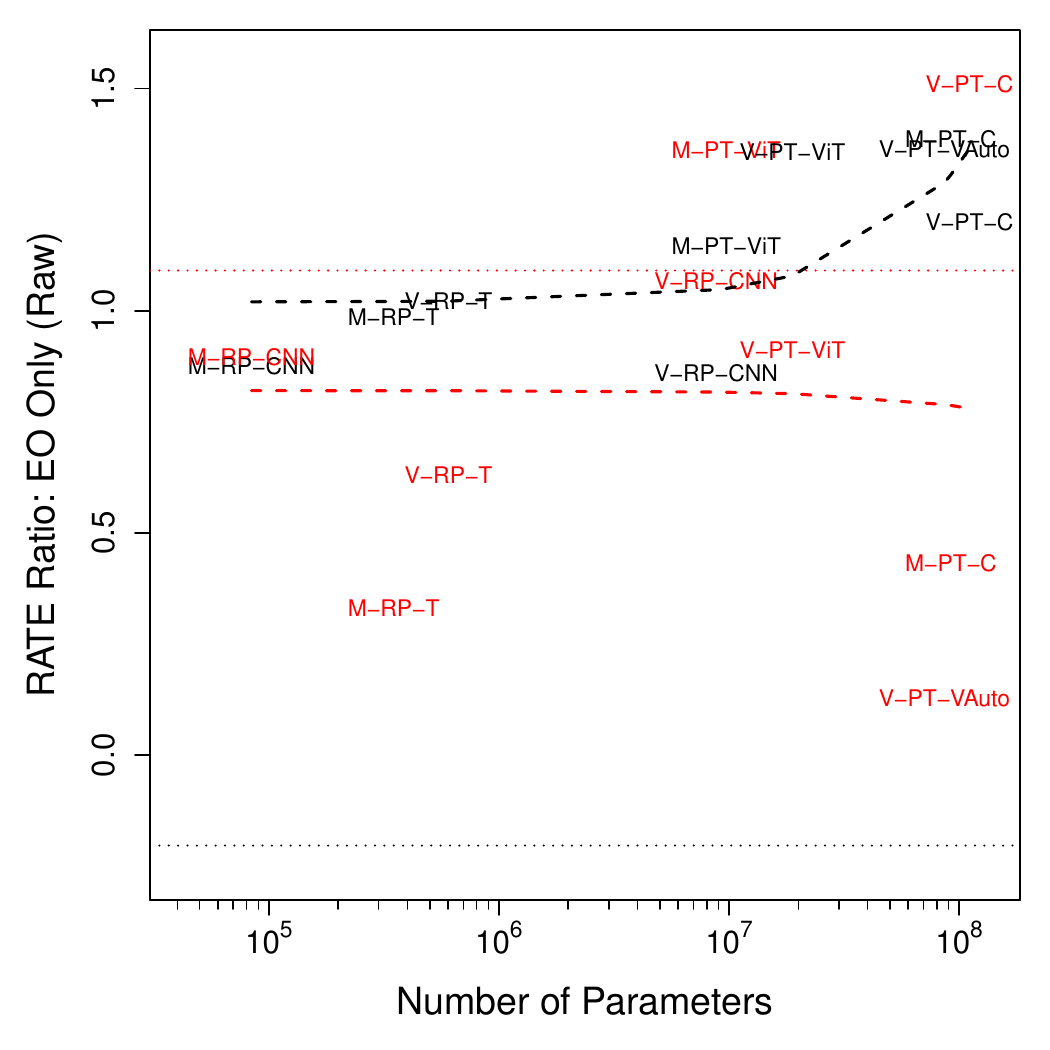}
\caption{
RATE ratios using \RATEWT{} weighting, plotted against number of image model parameters. Lower values along the $Y$ axis indicate weaker heterogeneity signal; higher values indicate stronger. ``M''/``V'' denote image/image sequence input data, respectively;  ``ViT'' denotes pre-trained Vision transformer; ``RP'' denotes random projection; ``C'' denotes Clay model. For results with QINI weighting, see Figure \ref{fig:SortAnalysisQINI}. Red denotes points from the Georgia analysis; black from Peru. Dashed lines indicating local averages are computed using splines; dotted red/black lines indicate RATE ratios for the tabular-only baseline. 
}
\label{fig:SortAnalysis}
\end{center}
\end{figure}

Figure \ref{fig:SortAnalysis} displays results from a RATE analysis where we examine the RATE ratio (ratio between the mean estimate and standard error) using \RATEWT{} weighting computed using cross-fitting. Higher values indicate a stronger heterogeneity signal under the model and data combination. We find that the Peru experiment shows a stronger average signal. The analysis shows that as the number of parameters used in the image or image sequence representation model grows, the RATE ratio also increases (and does not seem to flatten off, given the models available). Finally, we see an increase in the average RATE ratio for the EO sequence over image-only analyses. (See Figure \ref{fig:Robustness} for robustness analysis comparing R learner and causal forest results). 

\begin{figure}[ht!] 
  \begin{center}
\includegraphics[width=0.45\linewidth]{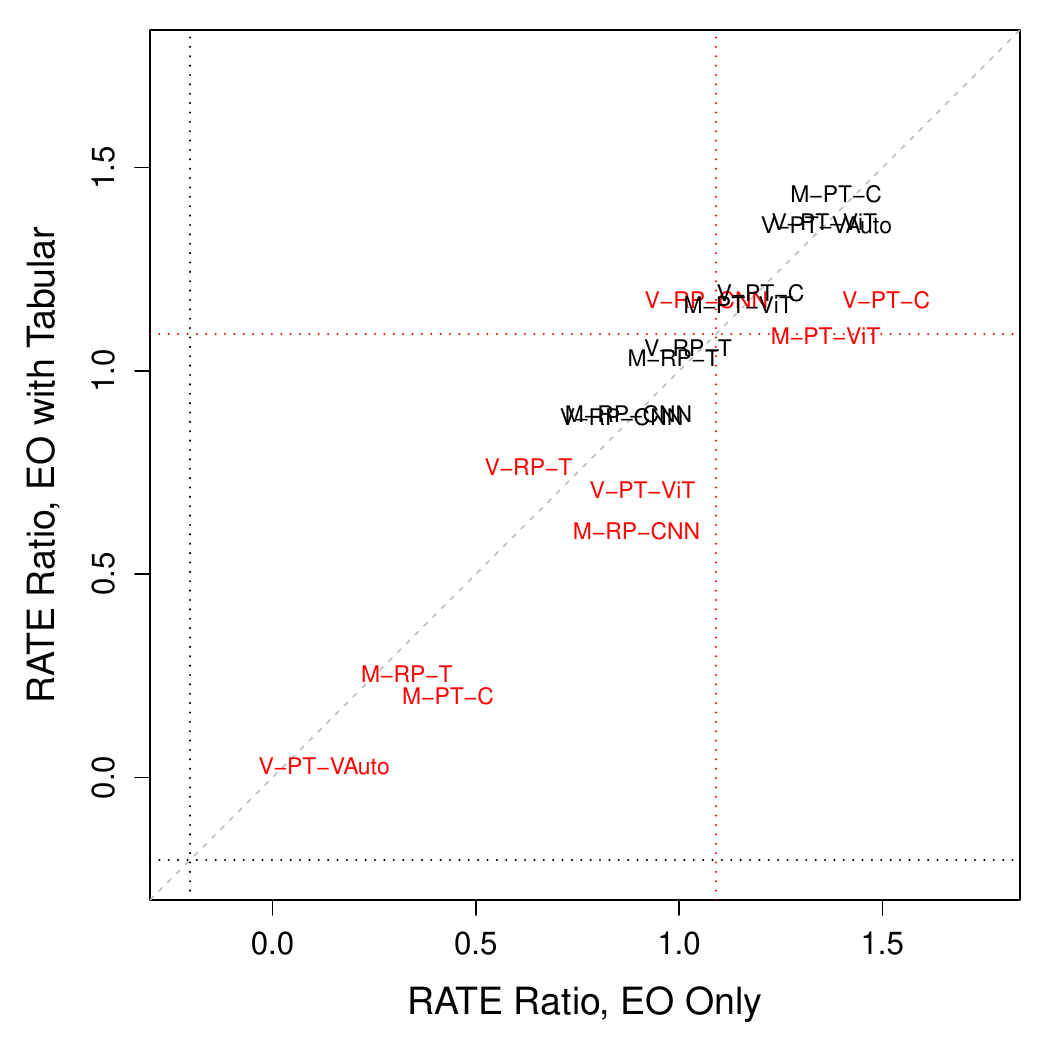}
\includegraphics[width=0.45\linewidth]{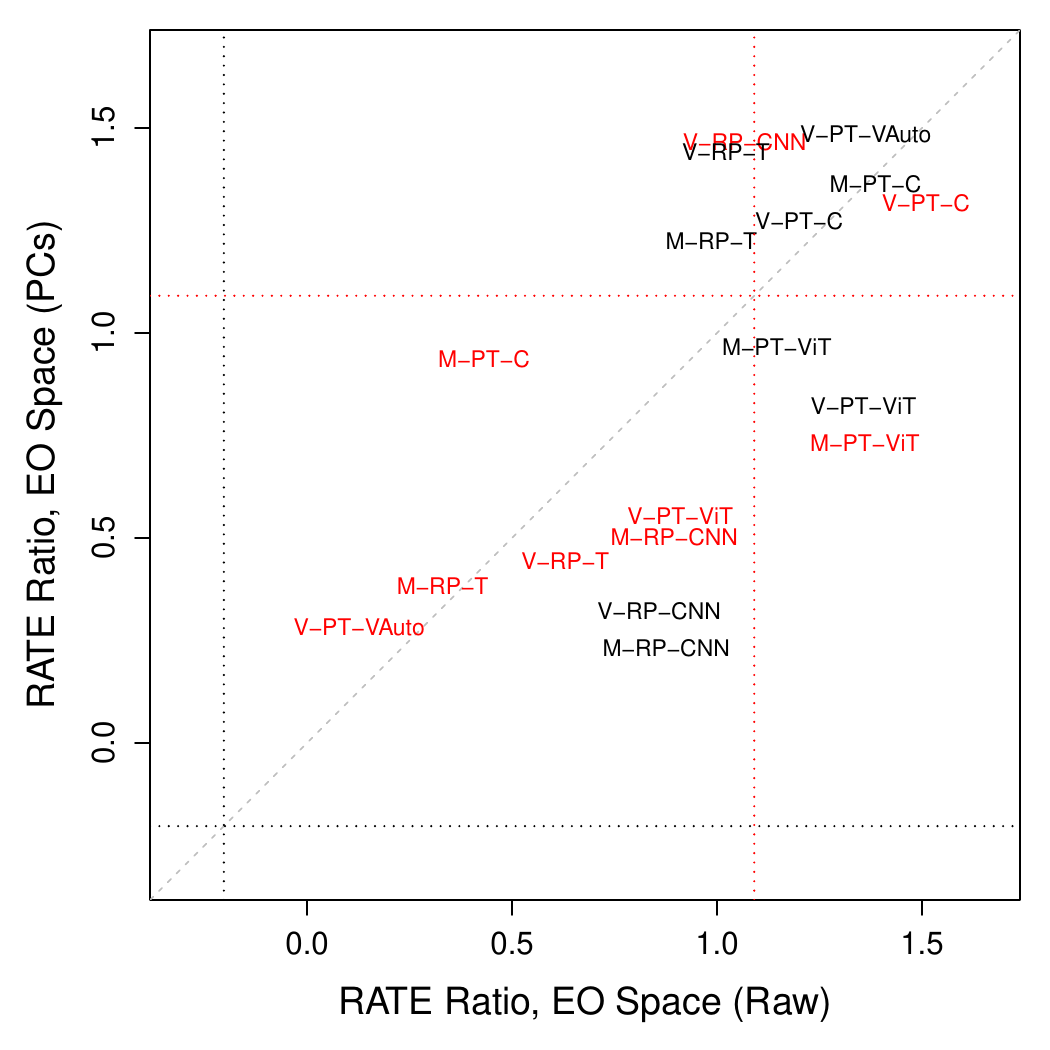}  
\caption{
  {\sc Left.} RATE ratios using \RATEWT{} weighting, EO only vs. EO with tabular. 
  {\sc Right.} RATE ratios, raw image sequence representation vs. principal component (PC) space. 
  See Figure \ref{fig:RawPCAnalysis\RATEWTALT} for results with \RATEWTALT{} weighting. ``M''/``V'' denote image/image sequence input data, respectively;  ``ViT'' denotes pre-trained Vision transformer; ``RP'' denotes random projection; ``C'' denotes Clay model. For results with QINI weighting, see Figure \ref{fig:SortAnalysisQINI}. Red denotes points from the Georgia analysis; black from Peru. Here, the gray dashed lines indicate the 45-degree line; dotted red/black lines indicate RATE ratios for the tabular-only baseline. 
}
\label{fig:RawPCAnalysis}
\end{center}
\end{figure}

It is worthwhile to dwell for a moment on the baselines presented in Figure \ref{fig:SortAnalysis}. The dotted red/black lines indicate the baseline RATEs in the tabular-only case for Georgia and Cusco, Peru. All EO RATEs exceed the Peru tabular RATE in the Peru case; in the Georgia case, a number of EO RATE ratios nearly reach the tabular values (see also Figure \ref{fig:RawPCAnalysis}). Let us pause on this finding: here, in these varied experimental settings, freely available satellite image sequences are able to provide a nearly as strong or stronger heterogeneity signal than when using pre-treatment covariates collected by researchers at great expense. Even where the RATE ratios are lower for EO-based inference, there is still value in the EO-based approach in that the tabular covariates are very often not available out of the study area. These two considerations should serve as motivation for the integration of EO data into RCT analysis going forward. 

Next, Figure \ref{fig:RawPCAnalysis\RATEWTALT} (left panel) analyses the relative gain of heterogeneity information when including tabular data in the heterogeneity analysis alongside the EO arrays. For this exercise, we directly append the tabular pre-treatment data variables to the image representations so analyzed. We find that, in the Georgia case, appendage of the tabular inputs to the EO representations seems to increase the RATE ratio, indicating improved heterogeneity signal. In the Peru case, the CATE signal is nearly unchanged with the inclusion of tabular inputs. 

The right panel of Figure \ref{fig:RawPCAnalysis\RATEWTALT} then analyzes the relationship between the heterogeneity signal in the raw and principal components space (10 dimensions). We find that, in the Georgia case, there may be some benefit in generating an improved heterogeneity signal, although in the Peru case, there may be a slight degradation. Overall, this exercise shows that the appendage of tabular or variable reduction through principal components does not dramatically improve heterogeneity estimation. (We analyze robustness to CATE model specification in Appendix II.)

\clearpage 
\subsection{Comparing Causal Estimates: EO Arrays vs. Land Cover Maps}

To further understand the information content contained in the heterogeneous treatment effects estimated from satellite image sequences, we now compare them to treatment effect estimates derived from detailed land cover maps of the experimental areas. Land cover maps are usually themselves derived from satellite data. They provide fine-grained information about the composition of the land surface, such as the extent of developed areas, agricultural land, forests, and water bodies. Changes in these land cover types over time reflect the evolution of the built and natural environment using constructed categories relevant to the environmental sciences. 

We probe whether summaries of these land cover categories provide similar or different information relative to the raw satellite images and image sequences in the study of causal effects. Strong agreement between the image-derived and land cover-derived estimates would provide convergent evidence for the ability of satellite image sequences to capture meaningful variation in treatment effects.

We first explore this relationship averaging across the various model representations used above. The protocol for generating land cover summaries is described in Appendix III. In brief, we take the proportion of land cover categories around each location as predictors in the effect heterogeneity model. Table \ref{tab:LandCorsTab_Peru_video_\RATEWT} and \ref{tab:LandCorsTab_Georgia_video_\RATEWT} the average correlation between CATEs from image sequence and land cover analysis, as well as the land cover RATE ratios. We see relatively small positive correlation; average values are higher in the Peru case than in the Georgia case, hinting at a stronger correspondence between the heterogeneity dynamic and the classification categories present in the land cover maps. Overall, this analysis suggests that the information contained in land cover representations related to effect heterogeneity seems to be relatively distinct from that contained in raw satellite image sequences.

\begin{table}[!htbp] \centering 
  \caption{Land cover comparisions, average correlation and RATE ratio (mean/std),  
                                                                average taken across the set of image sequence modeling strategies. 
                                                                Peru, video analysis. AUTOC weighting method in RATE calculation.
                                                               } 
  \label{tab:LandCorsTab_Peru_video_AUTOC} 
\begin{tabular}{@{\extracolsep{5pt}} ccc} 
\\[-1.8ex]\hline 
\hline \\[-1.8ex] 
 & Raw Space & PCs \\ 
\hline \\[-1.8ex] 
Correlation & $0.219$ & $0.211$ \\ 
RATE Ratio & $0.443$ & $0.486$ \\ 
\hline \\[-1.8ex] 
\end{tabular} 
\end{table}

\begin{table}[!htbp] \centering 
  \caption{Land cover comparisions, average correlation and RATE ratio (mean/std),  
                                                                average taken across the set of image sequence modeling strategies. 
                                                                Georgia, video analysis. AUTOC weighting method in RATE calculation.
                                                               } 
  \label{tab:LandCorsTab_Georgia_video_AUTOC} 
\begin{tabular}{@{\extracolsep{5pt}} ccc} 
\\[-1.8ex]\hline 
\hline \\[-1.8ex] 
 & Raw Space & PCs \\ 
\hline \\[-1.8ex] 
Correlation & $0.130$ & $0.089$ \\ 
RATE Ratio & $0.486$ & $0.043$ \\ 
\hline \\[-1.8ex] 
\end{tabular} 
\end{table}

Although the correlation between land cover and raw satellite estimated effects is relatively low, this does not necessarily answer the question of which of the two approaches should, in practice, be preferred. Answering this question is challenging for a number of reasons, particularly the unobservability of CATEs. Nevertheless, we can get some leverage into this question using another RATE analysis. 

\begin{figure}[ht!] 
 \begin{center}  
\includegraphics[width=0.45\linewidth]{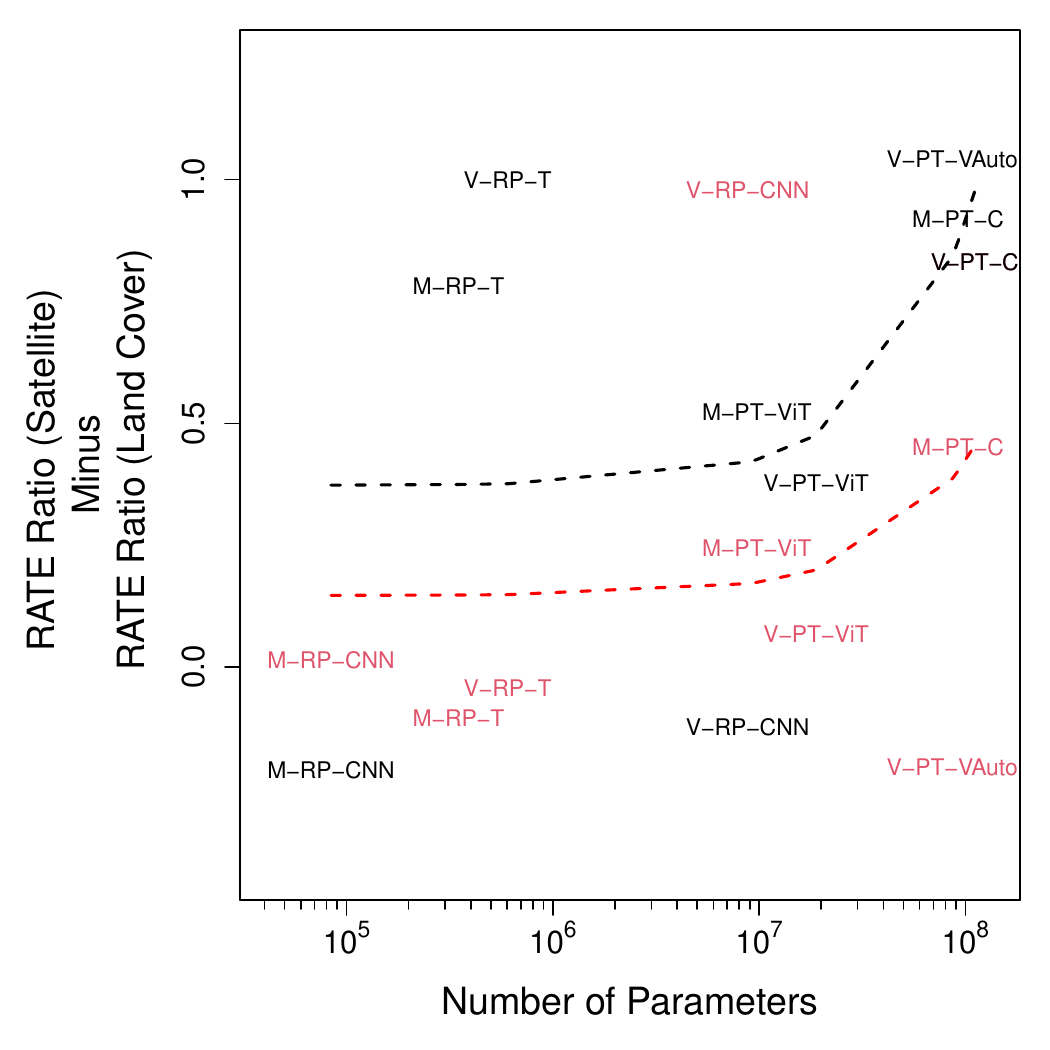}  
\caption{Lift from use of raw satellite imagery, as quantified by difference between the RATE measure from satellite analysis and from land cover analysis (positive values indicate relative benefit of raw EO representation over satellite-derived land cover representation). This analysis uses \RATEWT{} weighting in the RATE calculation; for analysis using \RATEWTALT{} weighting, see Figure \ref{fig:LandcoverRATE\RATEWTALT}.``M'' and ``V'' denote image and image sequence input data, respectively;  ``ViT'' denotes pre-trained Vision transformer; ``RP'' denotes random projection; ``C'' denotes Clay model. Dashed lines indicating local averages are computed using splines.
}
\label{fig:LandcoverRATE\RATEWT}
\end{center}
\end{figure}

Figure \ref{fig:LandcoverRATE\RATEWT} next shows the relative increase in RATE using raw image vs. land cover representations. In particular, we take: 
\begin{align*} 
\textrm{RATE Ratio}_{\textrm{EO}} - \textrm{RATE Ratio}_{\textrm{Land cover}},
\end{align*}
for the various EO data and model combinations. We find that, in general, using video representations with larger ML representation backbones yields a higher RATE from EO representations over land cover only. This suggests that, if the model class compute resources are not available, practitioners could consider using land cover as a post-processed EO representation in CATE estimation. 

This comparison against an alternative measurement of the experimental context provides greater contextualization for the use of image sequences to study treatment effect heterogeneity. By way of conclusion, we also note that land cover data are often updated only once every ten years. By contrast, satellite images have a weekly or even daily update frequency. Hence, investigators should consider satellite image sequences, especially in the study of rapidly moving signals that could be informative in some contexts, such as conflict or market dynamics.

\clearpage

\subsection{Multivariate Analysis}

The results provided so far examine just one or two aspects of data and modeling choices at a time as they affect heterogeneity signal. We now present multivariate results that provide a summary of our previous analyses and suggest insights for practitioners seeking to maximize heterogeneity signals in causal inference studies using satellite imagery. The outcome variable in these models, presented in Table \ref{tab:RegRATERatioRegs_SEanalytical} is the RATE ratio, which, as discussed, quantifies the strength of detected heterogeneity.

\begin{table}[htbp] \centering 
\footnotesize 
\begin{tabular}{@{\extracolsep{5pt}} lcc} 
\\[-1.8ex]\hline 
\hline \\[-1.8ex] 
 & Model 1 & Model 2 \\ 
\hline \\[-1.8ex] 
Video input (Baseline: Image input) & 0.10 (0.08) & 0.23 (0.07)$^*$  \\ 
Transformer (Baseline: CNN) & 0.07 (0.11) & 0.04 (0.10) \\ 
With tabular (Baseline: No tabular) & -0.06 (0.08) & -0.06 (0.07) \\ 
PCs (Baseline: No PCs) & -0.12 (0.09) & -0.12 (0.09) \\ 
Qini (Baseline: AUTOC) & -0.18 (0.07)$^*$  & -0.18 (0.07)$^*$  \\ 
Peru (Baseline: Georgia) & 0.32 (0.07)$^*$  & 0.32 (0.07)$^*$  \\ 
  &  &  \\ 
log(Parameter number) & 0.03 (0.02)$^*$  &  \\ 
Model - Random proj (Baseline: Clay) &  & -0.28 (0.09)$^*$  \\ 
--- PT-Video Autoencoder &  & -0.48 (0.15)$^*$  \\ 
--- PT-ViT &  & -0.07 (0.10) \\ 
  &  &  \\ 
\emph{Other statistics} &  &  \\ 
Observations & 108 & 108 \\ 
Adjusted R-squared & 0.23 & 0.31 \\ 
\hline \\[-1.8ex] 
\end{tabular} 
  \caption{Outcome: RATE ratio. Estimator: OLS.
                    Robust standard errors in parentheses.
                    $*$ denotes $p$ < 0.05. } 
  \label{tab:RegRATERatioRegs_SEanalytical} 
\end{table}

In Model 1 and 2 of Table \ref{tab:RegRATERatioRegs_SEanalytical}, we see results indicating that using video (image sequences) instead of static images increases the RATE ratio, suggesting that temporal information improves our ability to detect heterogeneity. Regarding evaluation metrics, QINI weighting shows a lower RATE ratio compared to AUTOC, indicating that AUTOC may be more sensitive to heterogeneity in our contexts (perhaps due to heterogeneity restricted to a small subset of units). Notably, the Peru experiment shows a higher RATE ratio compared to Georgia holding all else equal, suggesting that the local context significantly influences our ability to detect heterogeneity. The effects of QINI weighting and experimental context remain consistent. 

In terms of model architecture, random projection and non-EO pre-trained models show a decrease in RATE ratio compared to the Clay foundation model (see Model 2) This suggests that domain-specific pre-training may be beneficial for capturing relevant heterogeneity. Model 1 simplifies the analysis by focusing on the number of model parameters. The results reveal a positive relationship between the log number of parameters and the RATE ratio, suggesting that more complex models may be better at capturing heterogeneity. 

These findings offer several practical implications. First, the use of image sequences, when possible, is recommended, as they provide richer information for detecting heterogeneity. Second, researchers should consider using more complex models with a larger number of parameters trained on domain-specific inputs, as they tend to yield higher RATE ratios. Third, AUTOC weighting is preferable to QINI for potentially stronger heterogeneity signals. Fourth, researchers should be aware that the ability to detect heterogeneity may vary significantly across different geographic or experimental contexts. Lastly, inclusion of tabular data or use of principal components show no improvements in detecting heterogeneity.


\subsection{Transportability Analysis}

One strength of the EO-based causal inference lies in the availability of satellite data outside the experimental context. Many covariates, like age and gender, are most often only accessible for units in the experiment, limiting the ability to use knowledge of heterogeneity from those features in areas outside the original study. EO data, however, is available both within and outside the experimental study. Additionally, tabular CATEs may contain different statistical information than the EO image sequence CATEs, as shown in Figure \ref{fig:CorTabEO}. Moreover, in some situations, such as in the allocation of economic aid,  interventions are done at a level of spatial aggregation (e.g., village level), indicating the policy relevance of image sequence CATEs based on EO arrays.

The promise of experimental transportability based on satellite images still requires some assumptions. First, for the sake of simplicity, we assume no spillovers in the CATE dynamics. In other words, conditional on the EO array, each location will have a distinct treatment effect that is not influenced by the treatment allocation of other nearby units. Second, we also assume the correct specification of the heterogeneity model. Hence, our models are extrapolating from experimental to non-experimental sites using the underlying heterogeneity model (see Section \ref{s:Conclusion} for discussion). Our quantity of interest is thus:  
\begin{equation}
\tau(\bv)_{\textrm{Transport}} = \mathbb{E}\left[ Y_{i}(1) - Y_{i}(0) \mid \bV_{i} = \bv, S_i = 0\right], 
\end{equation}
for units $i$ outside the experimental context (for whom the experimental selection indicator $S_i$ is 0).

\begin{figure}[ht!] 
\begin{center}   \includegraphics[width=0.95\linewidth]{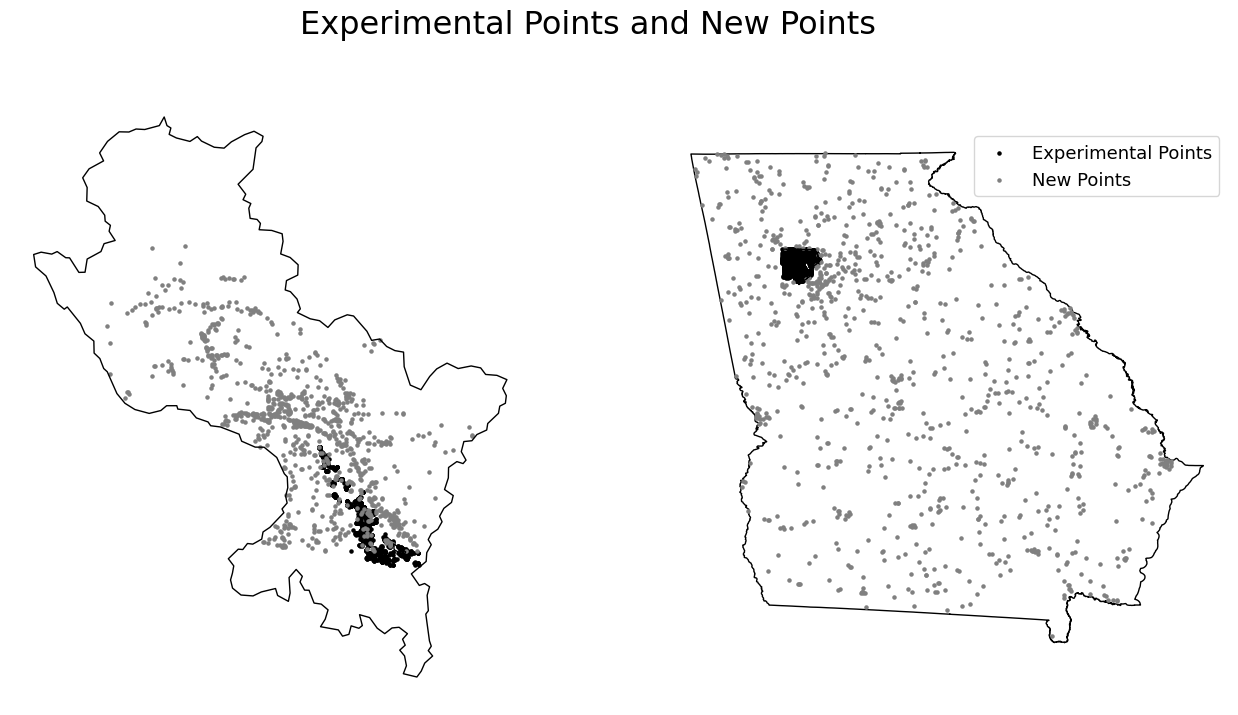}
  \caption{
{\sc Left.} Cusco transportability points (gray); experimental points (black). Due to extreme population sparsity in the northern region, transportability points sampled via population weighting. 
  {\sc Right.} Georgia transportability points. Transportability points sampled via population weighting. Transportability points sampled uniformly at random. 
}
\label{fig:TransportDescriptives}
\end{center}
\end{figure}

With these assumptions in mind, we generate transportability effect maps for all of Georgia in the US and the entire Cusco region in Peru as follows. First, we generate additional locations within the state of Georgia to do transportability analysis using the trained EO heterogeneity models. We obtain the locations of 1000 census blocks outside Cobb County and their corresponding satellite image sequences over the same time period considered for the experimental data. Likewise, we also generate additional locations within the Cusco region of southeastern Peru using the same approach as Georgia in the US. Here, however, we sample new locations with population weights, since much of northern Cusco is uninhabited. Population weighting enables more precise spatial estimates for the regions of policy interest. Figure \ref{fig:TransportDescriptives} visualizes this sampling process.

\begin{figure}[ht!] 
  \begin{center}
\includegraphics[width=0.45\linewidth]{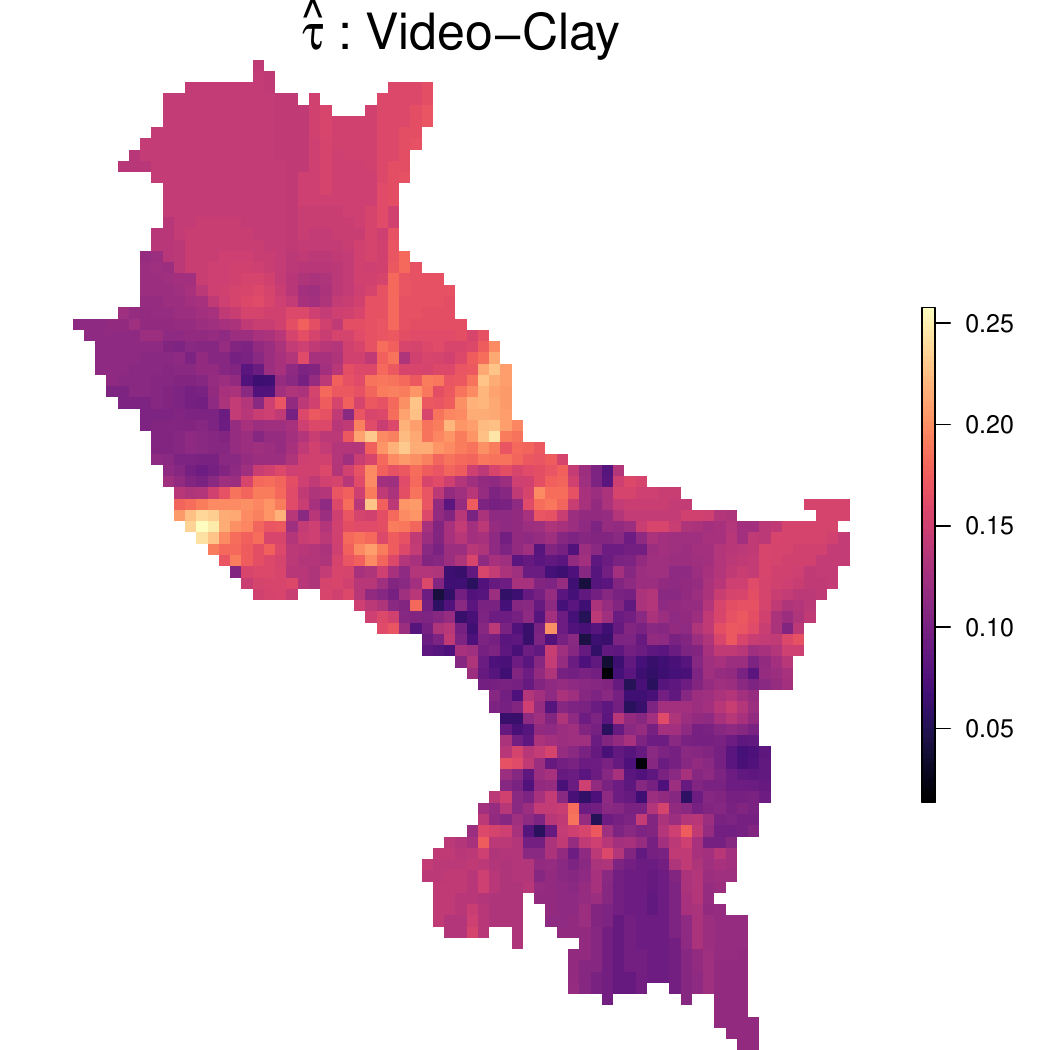}
\includegraphics[width=0.45\linewidth]{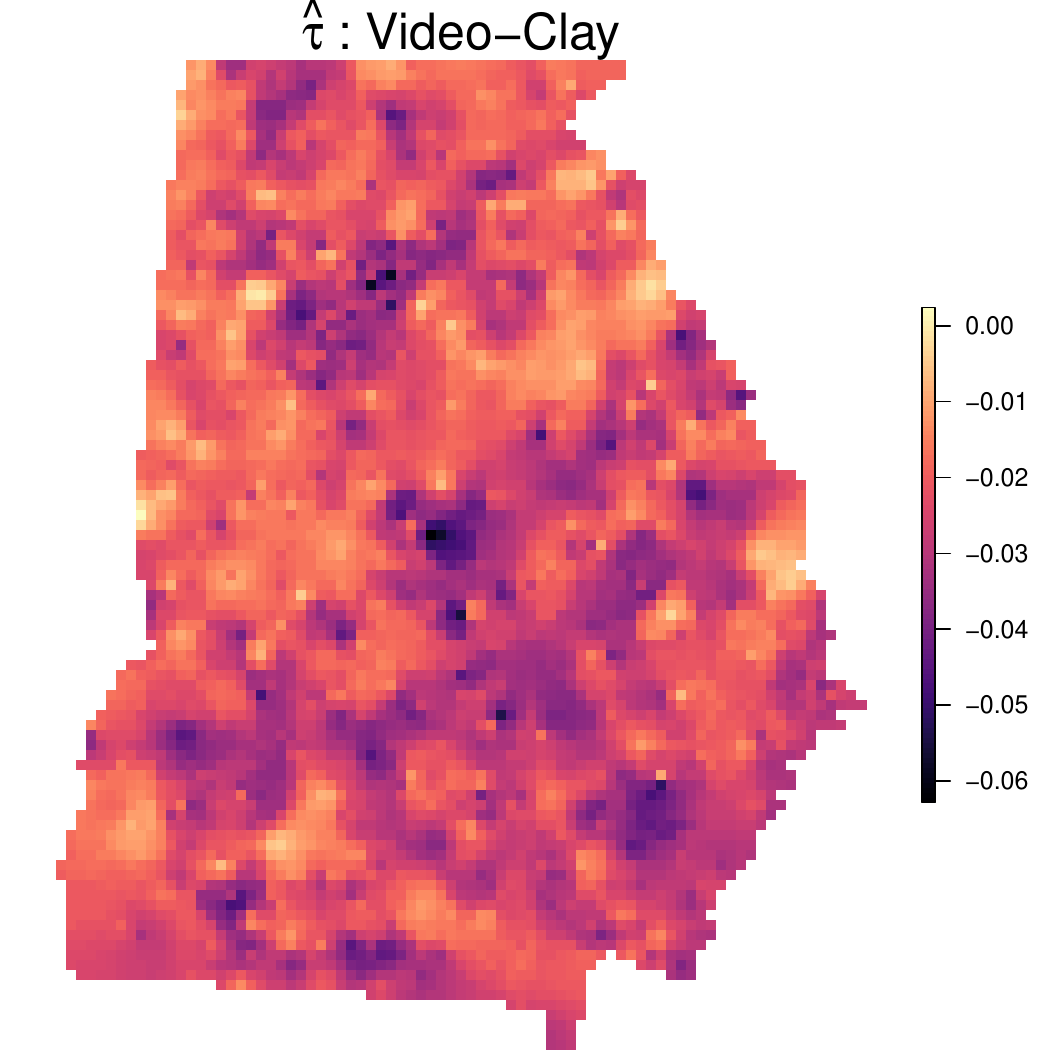}
\caption{
Transportability analysis using image sequence data in Cusco {\sc (Left)} and Georgia {\sc (Right)}. Colors indicate treatment effect estimate values under the model specification (yellow $>$ blue). See Figures \ref{fig:TransportClayImage} and \ref{fig:TransportVideoMAE} for transportability maps using the Clay model with image inputs only and the Masked Video Autoencoder. 
}
\label{fig:Transport}
\end{center}
\end{figure}

We first examine the transportability map from CATEs constructed using the Clay model representations. Figure \ref{fig:Transport} visualizes the estimated CATEs for all parts of Georgia and Cusco. We see in the Georgia case a few areas have positive treatment effects (i.e., increased water consumption in response to the conservation intervention), with negative treatment effects, especially distributed towards the center of the state. In the Peru case, the results are spatially heterogeneous, with the southern, more populous regions displaying consistently smaller effects compared to the more remote mountainous areas of central Cusco. Appendix II contains heterogeneity maps using the other modeling strategies in map construction; the pattern just described is generally consistent across the various modeling strategies used, with some small variations. 

\begin{figure}[ht!] 
 \begin{center}  
\includegraphics[width=0.90\linewidth]{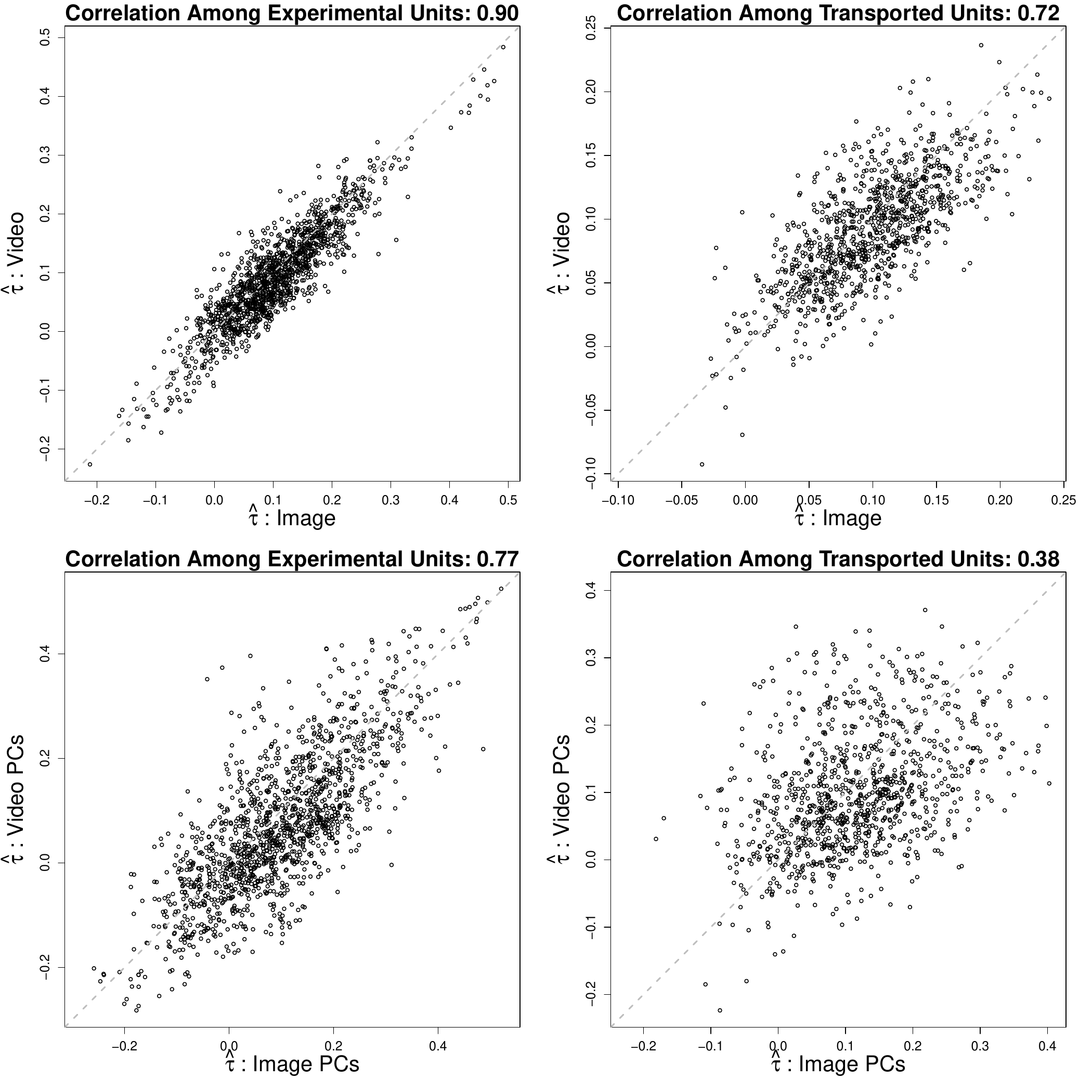}
\caption{
Peru transportability analysis. {\sc Top/Bottom.} Relationship between image and image sequence CATEs in the raw representation space and principal component space. {\sc  Left/Right.} Relationship among experimental units and units used in transportability analysis. 
}
\label{fig:PeruCorsTransport}
\end{center}
\end{figure}

\begin{figure}[ht!] 
 \begin{center}  
\includegraphics[width=0.90\linewidth]{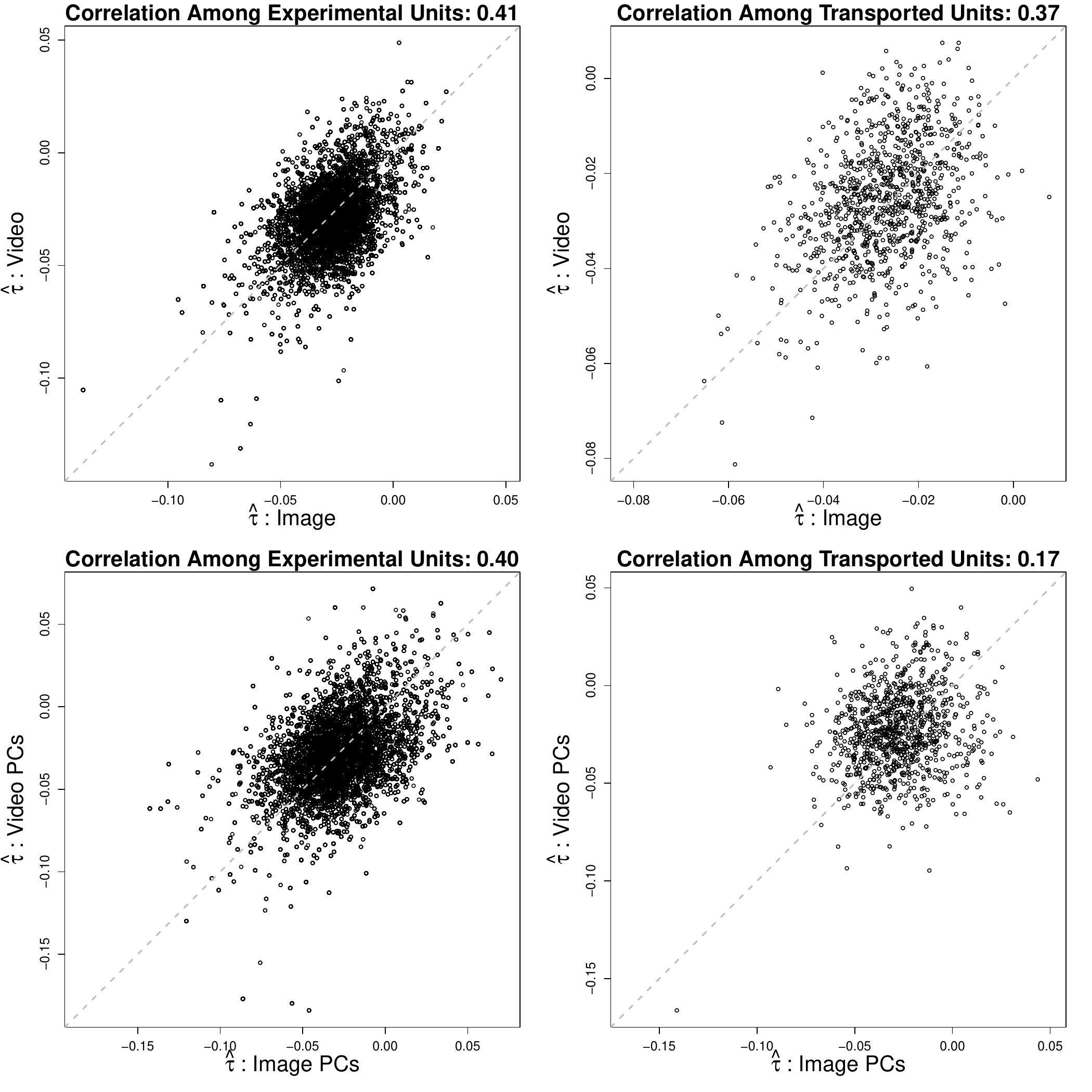}
\caption{
Georgia transportability analysis. {\sc Top/Bottom.} Relationship between image and image sequence CATEs in the raw representation space and principal component space. {\sc  Left/Right.} Relationship among experimental units and units used in transportability analysis. 
}
\label{fig:GeorgiaCorsTransport}
\end{center}
\end{figure}

We next explore in Figures \ref{fig:PeruCorsTransport} and \ref{fig:GeorgiaCorsTransport} the relationship between the image and image sequence CATEs in the raw representation and principal component representation space, both among experimental and transported units. We see a stronger correlation between image and video representations in the raw space in both Peru and Georgia. We also find a reduced correlation among transported sites---indicating the heightened sensitivity of transportability construction to model and data used in their construction. We also find that the relationship between results with image and image sequence representations is weakened further in the transported sites when using the principal component representations (instead of the raw values). These findings indicate that, while the principal components may reduce the dimensionality of the high-dimensional arrays, this dimensionality reduction may reduce the robustness of transportability results to data choice. Whether this robustness is or is not desirable would be context dependent. 

\clearpage 
\section{Discussion \& Conclusion}\label{s:Conclusion}

In this paper, we make several contributions to the field of causal inference using satellite imagery. First, we develop a methodological framework for leveraging image sequence data to estimate heterogeneous treatment effects in randomized controlled trials. This approach allows researchers to capture rich historical and contextual information that may influence treatment response. 

Second, we demonstrate the utility of this framework through both simulations and applications to two real-world RCTs---an anti-poverty intervention in Peru and a climate adaptation study in Georgia, US. These analyses highlight how our methods can be applied across different scales and contexts; comparison of image sequence-based approaches to land cover maps provides insight into the specific value of raw satellite imagery for causal inference.

Finally, we explore how these techniques can enable improved transportability of experimental results to new geographic areas. By making our data and code open-source, we hope to facilitate further methodological development and empirical applications at the intersection of causal inference, computer vision, and earth observation. Ultimately, this work aims to enhance our ability to study and understand the causal impacts of interventions on complex socio-environmental systems across space and time.

The comparison between our image sequence-based approach and land cover map analysis reveals interesting insights. While there is some correlation between the two, the relatively low $R^2$ values between them suggest that raw satellite imagery captures aspects of local context and development trajectories that are not fully reflected in pre-defined land cover categories. This underscores the value of using high-dimensional, continuous measures of the built and natural environment in causal analysis, rather than relying exclusively on discretized classifications.

Our EO-ML approach for causal inference is not without limitations motivating future work. First, while our transportability analysis demonstrates the usefulness of using EO-ML models for generalizing experimental results, there are some challenges. The variability of our estimates across model specification and representation choices highlights the need for careful validation and robustness checks when applying these methods in practice. More research is required to identify procedures that reduce that variability in a systematic way. 

Second, our current approach relies on the assumption of no spillovers in the CATE dynamics. The transportability maps of Figure \ref{fig:Transport} present CATE estimates for the entirety of the geographic context in which the Cusco and Cobb County experiments took place. While these estimates might be useful in a policy exercise where only a few treatment allocations can be made, if a large number of interventions can be further performed, large-scale spillovers seem more likely (and has been empirically noted, e.g., in the context of economic aid \citep{demir2024target}). Future work should explore how to incorporate spillover effects into our framework, possibly by leveraging the spatial information inherent in satellite imagery to model inter-unit dependencies. 

Third, another area for future research is the integration of our approach with causal machine-learning methods designed for observational data \citep{daoud_melting_2021,chernozhukov2018double}. While we focused on experimental settings, many policy-relevant questions involve non-random interventions. Indeed, in many contexts, replication data from experiments do not contain information, even on units' approximate locations, in order to maintain the privacy of experimental subjects. This stands in contrast to many experimental studies (such as the analysis of economic aid), where there are precise geolocations specific to each intervention. Thus, the observational extension of the ideas presented here is worth future effort. 

Fourth, model and estimation uncertainty is an important consideration. Although there is an increasing interest in quantifying uncertainty in deep learning models \citep{kakooei_increasing_2024}, more statistical theory is required to tailor such estimation for CATE analysis with EO-ML models. A related question concerns whether temporal aggregation should occur before or after spatial aggregation (or jointly) when forming representations from satellite image sequences.

Future work can also relax assumptions used in the transportability analysis to address issues regarding selection bias. To what extent can an EO-ML model reveal selection bias on why some sites, but not others, were included in the experiment? With that bias estimation, one would, in principle, be able to adjust the resulting estimates via weighting \citep{westreich2017transportability}.

As EO technologies continue to advance, with higher resolution sensors and more frequent revisit times, the potential for satellite imagery to inform causal inference will only grow  \citep{burkeUsingSatelliteImagery2021,daoud_statistical_2023}. Developing scalable and interpretable methods to harness this emerging data source for addressing policy-relevant causal questions should garner further interest in the coming years. \hfill $\square$

\section{Data Availability Statement}
\noindent All methods are accessible in an open-source codebase available at
\begin{itemize}
\item[] \href{https://github.com/AIandGlobalDevelopmentLab/causalimages-software}{\url{GitHub.com/AIandGlobalDevelopmentLab/causalimages-software}}
\end{itemize}

Replication data (including satellite images and experimental data) for the Georgia experiment will be made available in a Harvard Dataverse:
\begin{itemize}
\item[] \href{https://doi.org/10.7910/DVN/MHWQTR}{\url{doi.org/10.7910/DVN/MHWQTR}}
\end{itemize} 
as well as a Hugging Face repository: 
\begin{itemize}
\item[]\href{https://huggingface.co/datasets/cjerzak/CausalImSeq}{\url{HuggingFace.co/datasets/cjerzak/CausalImSeq}}
\end{itemize}

\section{Acknowledgements}\label{s:Acknowledge}
We thank participants of the 2023 Interactive Causal Learning Conference and members of the AI \& Global Development Lab for helpful feedback. We thank Dean Karlan and Andre Nickow for assistance in accessing data for the Peru experiment. We thank Hannah Druckenmiller, Antonio Linero, and SayedMorteza Malaekeh for helpful discussions. We thank Fucheng Warren Zhu for truly excellent research support.  We thank Carlos Cinelli for guidance. 

\printbibliography

\clearpage \newpage 
\begin{center}
{\Huge \sc Supplementary Information} 
\end{center}

\renewcommand{\thefigure}{A.I.\arabic{figure}}
\setcounter{figure}{0}  

\renewcommand{\thetable}{A.I.\arabic{table}}
\setcounter{table}{0}  

\renewcommand{\thesection}{A.I.\arabic{section}}
\setcounter{section}{1}

\section*{Appendix I. Additional Simulation Results}
\begin{figure}[ht!]
 \begin{center}  
   \includegraphics[width=0.45\linewidth]{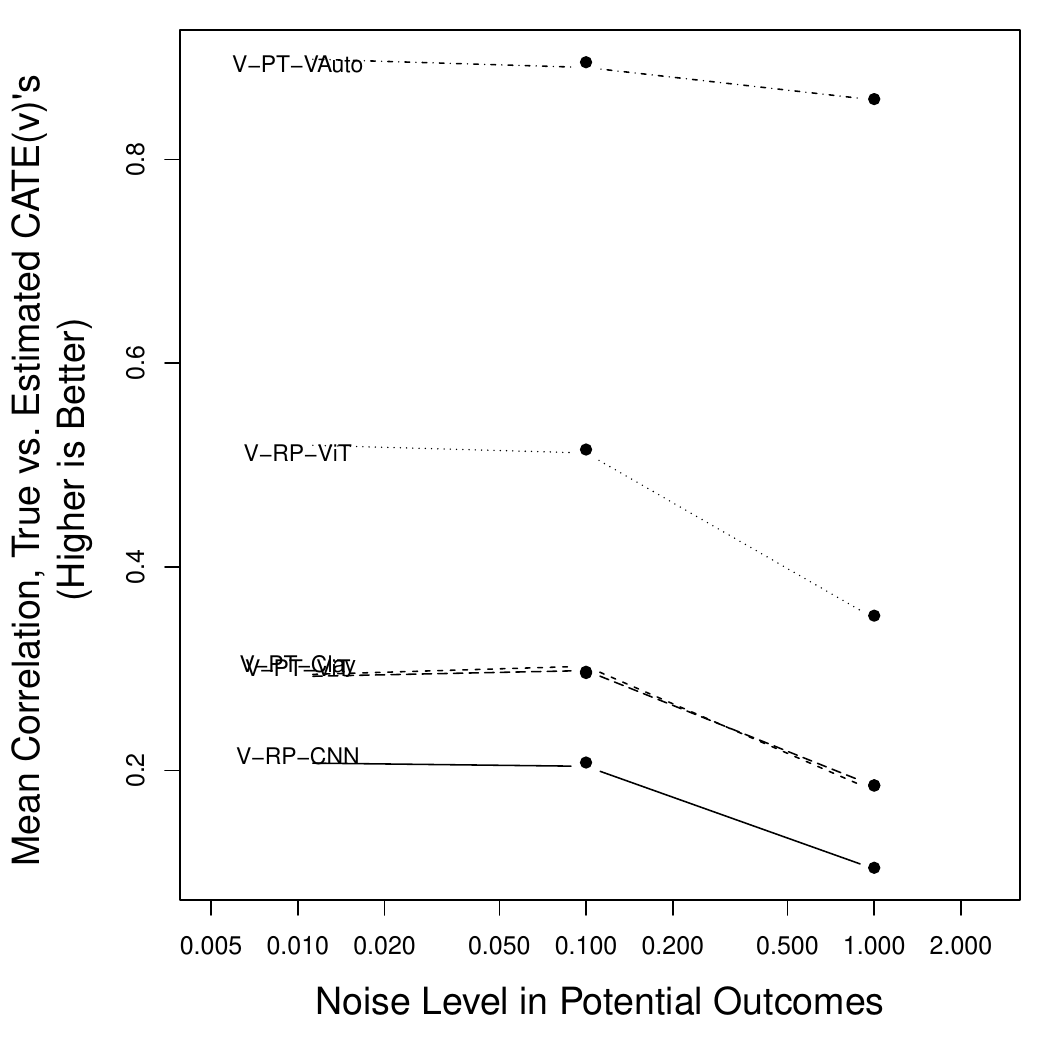}
   \includegraphics[width=0.45\linewidth]{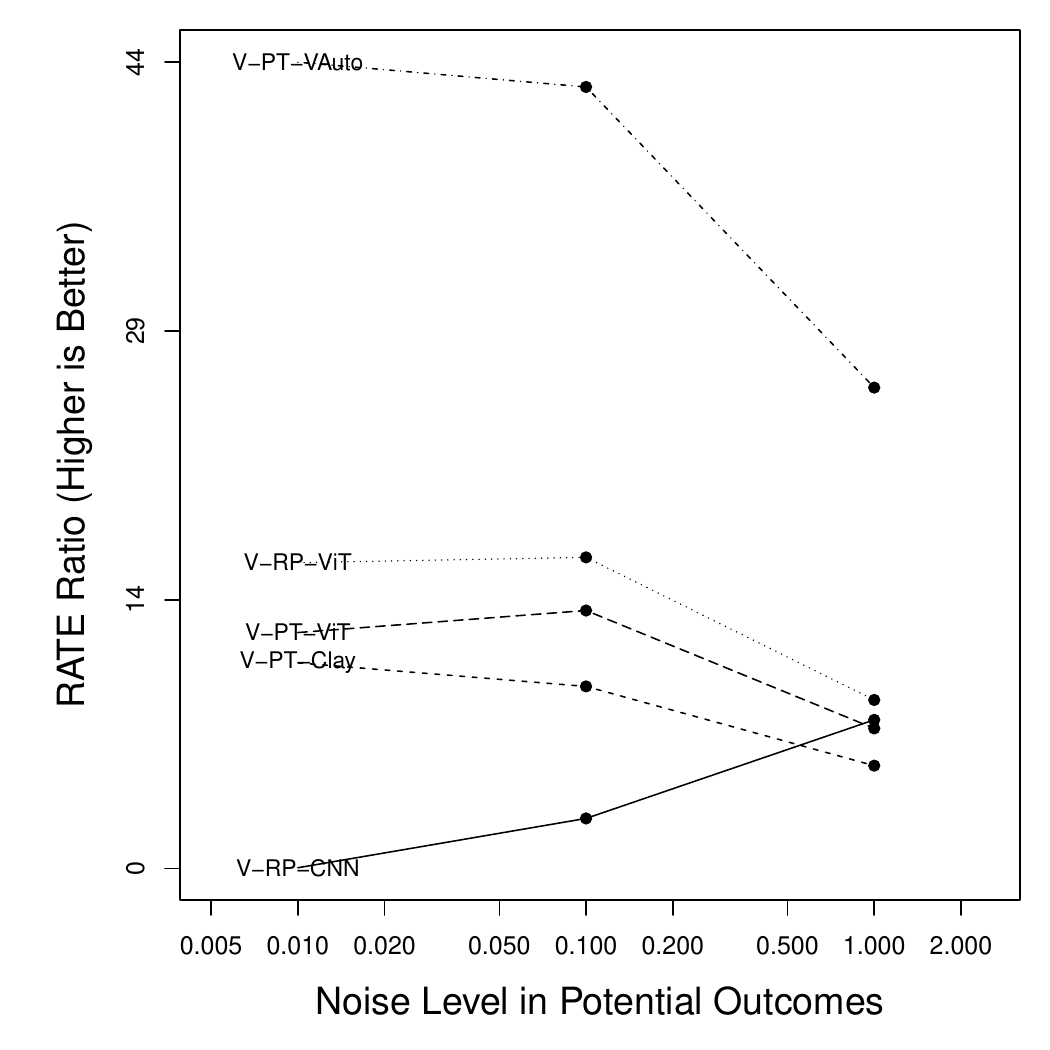}
   \caption{Simulation results, with principal component representations used in a causal forest heterogeneity model.}
\label{fig:SimMainPC}
\end{center}
\end{figure}

\begin{figure}[ht!]
 \begin{center}  
\includegraphics[width=0.5\linewidth]{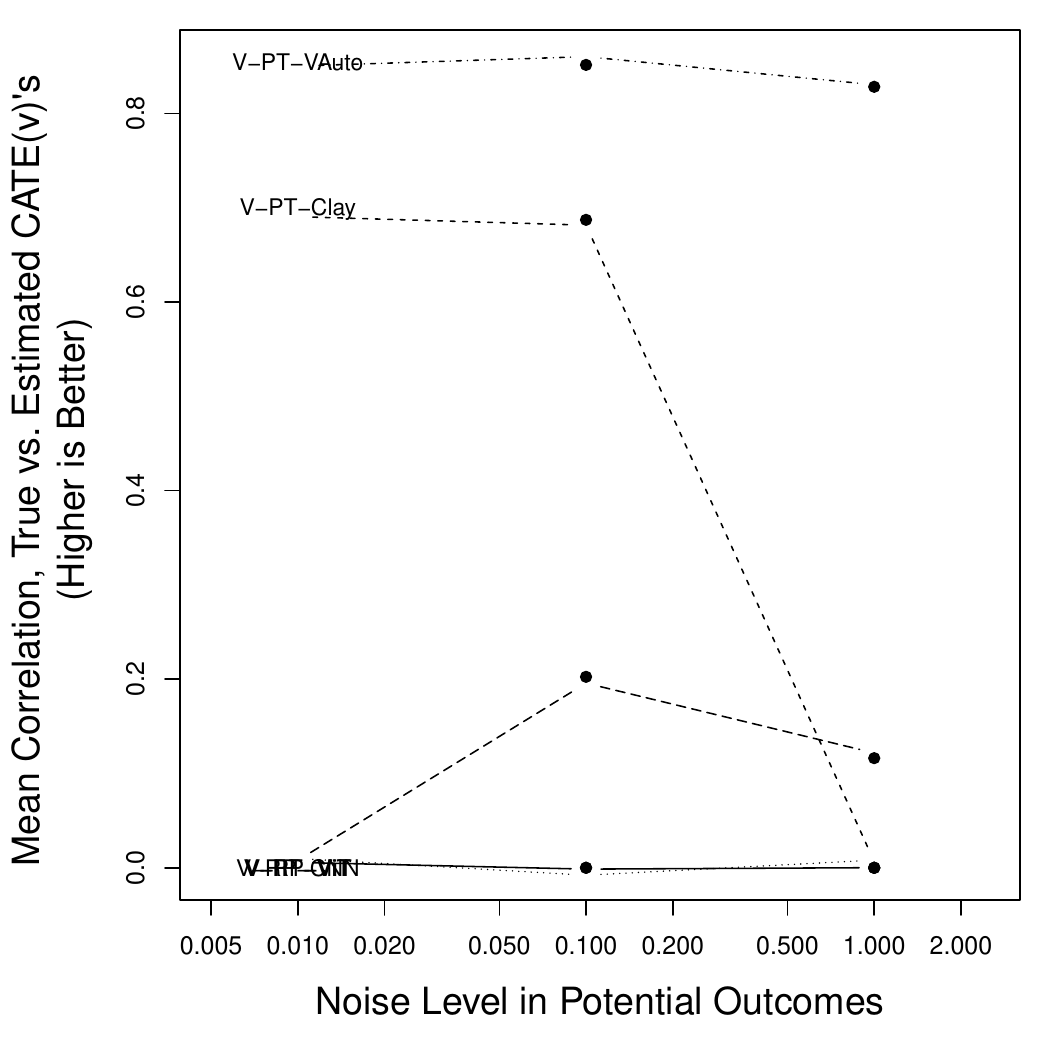}
   \caption{Simulation results, with a lasso-based R learner heterogeneity model. At the 0 correlation values, the lasso model generated no estimated heterogeneity. }
\label{fig:SimR}
\end{center}
\end{figure}

\begin{figure}[ht!]
 \begin{center}  
\includegraphics[width=0.5\linewidth]{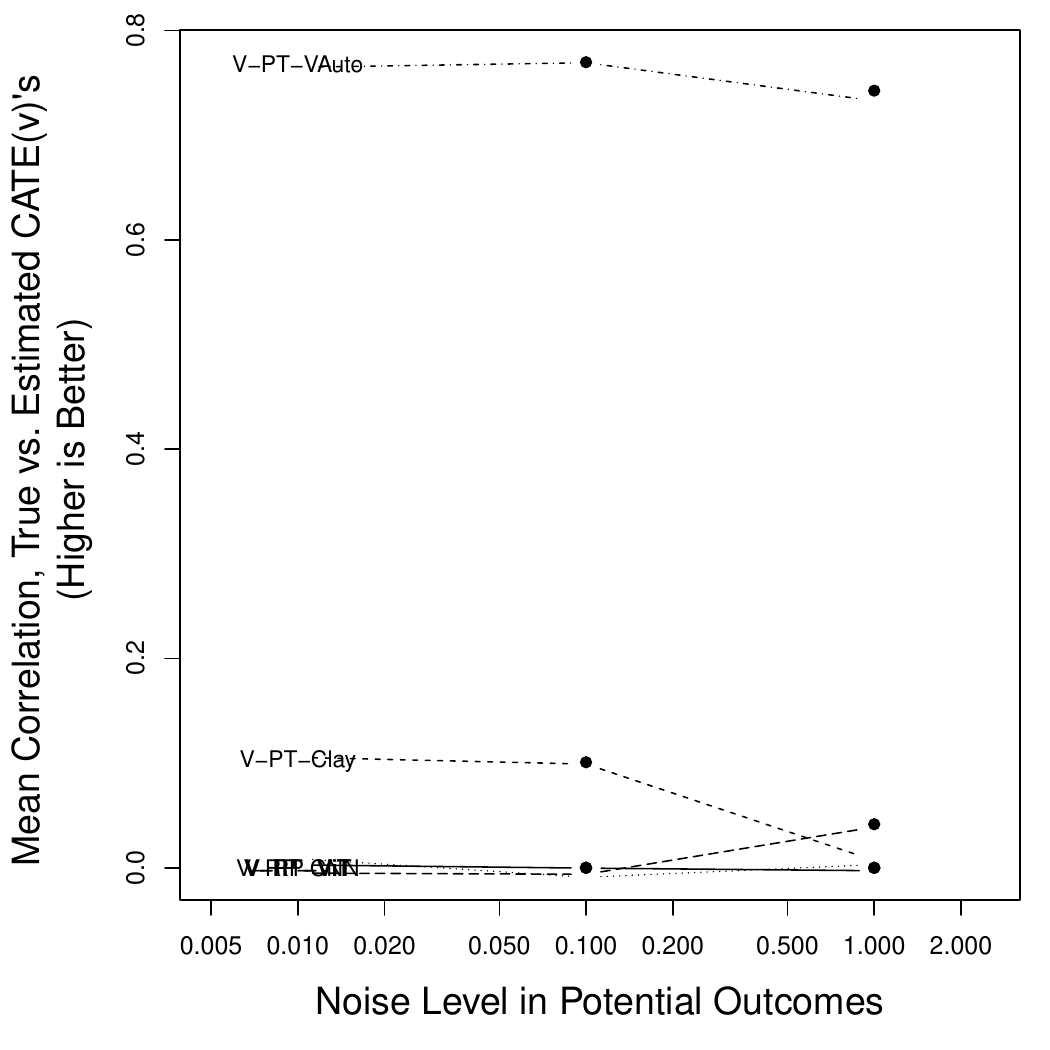}
   \caption{Simulation results, with principal component representations used in a lasso-based R learner heterogeneity model. At the 0 correlation values, the lasso model generated no estimated heterogeneity.
   }
\label{fig:SimRPC}
\end{center}
\end{figure}

\clearpage\newpage 
\renewcommand{\thefigure}{A.II.\arabic{figure}}
\setcounter{figure}{0}  

\renewcommand{\thetable}{A.II.\arabic{table}}
\setcounter{table}{0}  

\renewcommand{\thesection}{A.II.\arabic{section}}
\setcounter{section}{1}
\section*{Appendix II. Additional Empirical Analyses}

\begin{figure}[ht!] 
 \begin{center}  
\includegraphics[width=0.45\linewidth]{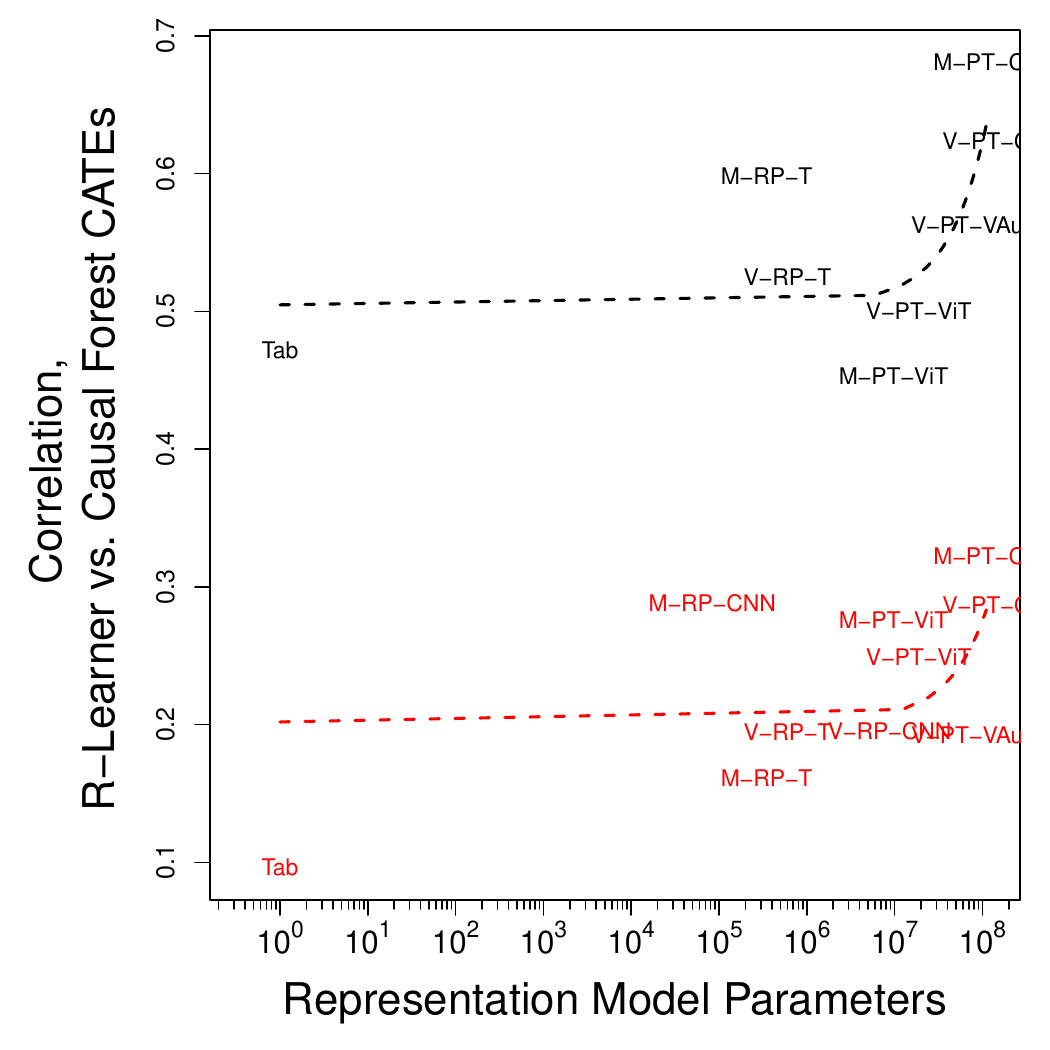}  
\caption{
  Correlation between estimated CATEs (R learner and causal forest) as a function of parameter number in the image sequence representation. Red represents points from the Georgia analysis; black from the Peru analysis. Dashed lines indicating local averages are computed using splines; dotted red/black lines indicate RATE ratios for the tabular-only baseline. 
}
\label{fig:Robustness}
\end{center}
\end{figure}

Figure \ref{fig:Robustness} shows that the correlation between CATEs under causal forest and R-learner (lasso-based) estimation is positive and much higher in the Peru case than in the Georgia case. This result could indicate the presence of more intricate non-linear heterogeneity relationships in the Georgia case that are not well-modeled linearly. 

\begin{figure}[ht!] 
 \begin{center}  
\includegraphics[width=0.55\linewidth]{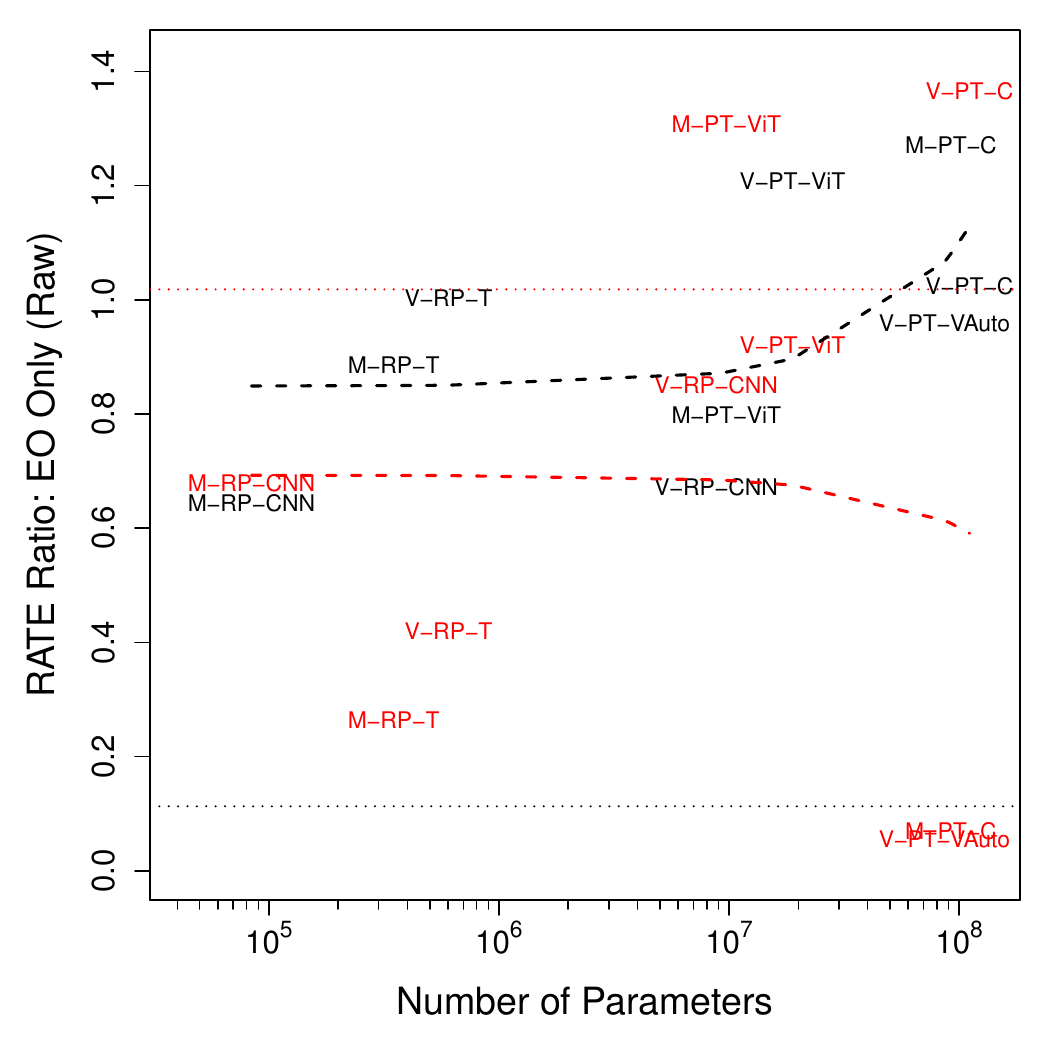}
\caption{
RATE ratios, \RATEWTALT{} weighting, plotted against number of image model parameters. Lower values along the $Y$ axis indicate weaker heterogeneity signal; higher values indicate stronger. Red points denote Georgia RCT results; black points denote Peru results; ``M''/``V'' denote image/image sequence input data, respectively; ``ViT'' denotes pre-trained Vision transformer; ``RP'' denotes random projection; ``C'' denotes Clay model. Dashed lines indicating local averages are computed using splines; dotted lines indicate the baseline RATEs from tabular-only analysis. 
}
\label{fig:SortAnalysis\RATEWTALT}
\end{center}
\end{figure}

\begin{figure}[ht!] 
  \begin{center}
\includegraphics[width=0.45\linewidth]{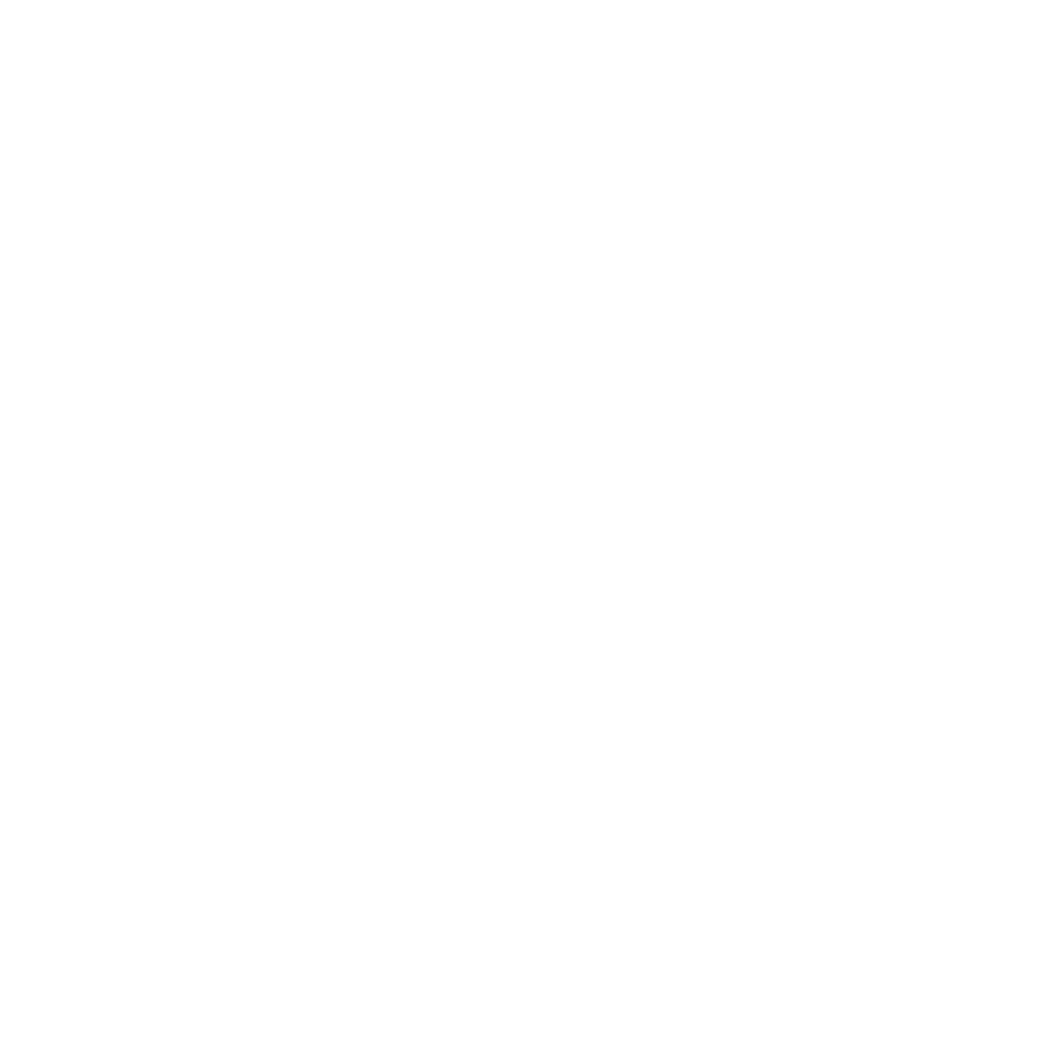}
\includegraphics[width=0.45\linewidth]{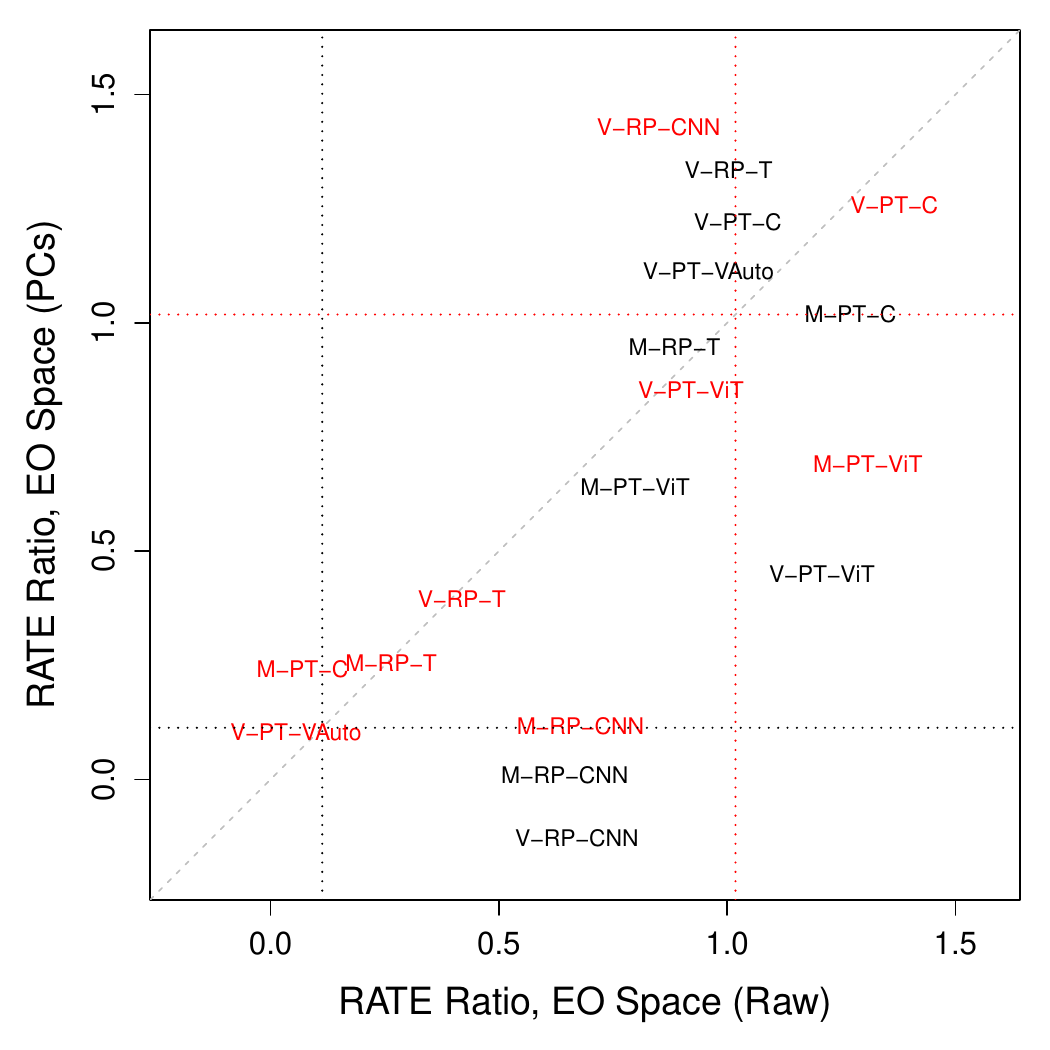}  
\caption{
  {\sc Left.} RATEs, tabular and vs. raw image sequence representation space. \RATEWTALT{} weighting in RATE measure.
  {\sc Right.} RATEs, raw image sequence representation vs. principal component space. \RATEWTALT{} weighting in RATE measure.
  Here, the gray dashed lines indicate the 45-degree line; dotted red/black lines indicate RATE ratios for the tabular-only baseline. 
}
\label{fig:RawPCAnalysisQINI}
\end{center}
\end{figure}

\begin{table}[!htbp] \centering 
  \caption{Land cover comparisions, average correlation and RATE ratio (mean/std),  
                                                                average taken across the set of image sequence modeling strategies. 
                                                                Peru, video analysis. QINI weighting method in RATE calculation.
                                                               } 
  \label{tab:LandCorsTab_Peru_video_QINI} 
\begin{tabular}{@{\extracolsep{5pt}} ccc} 
\\[-1.8ex]\hline 
\hline \\[-1.8ex] 
 & Raw Space & PCs \\ 
\hline \\[-1.8ex] 
Correlation & $0.219$ & $0.211$ \\ 
RATE Ratio & $0.094$ & $0.038$ \\ 
\hline \\[-1.8ex] 
\end{tabular} 
\end{table}

\begin{table}[!htbp] \centering 
  \caption{Land cover comparisions, average correlation and RATE ratio (mean/std),  
                                                                average taken across the set of image sequence modeling strategies. 
                                                                Georgia, video analysis. QINI weighting method in RATE calculation.
                                                               } 
  \label{tab:LandCorsTab_Georgia_video_QINI} 
\begin{tabular}{@{\extracolsep{5pt}} ccc} 
\\[-1.8ex]\hline 
\hline \\[-1.8ex] 
 & Raw Space & PCs \\ 
\hline \\[-1.8ex] 
Correlation & $0.130$ & $0.089$ \\ 
RATE Ratio & $0.095$ & $0.039$ \\ 
\hline \\[-1.8ex] 
\end{tabular} 
\end{table}

\begin{figure}[ht!] 
 \begin{center}  
\includegraphics[width=0.45\linewidth]{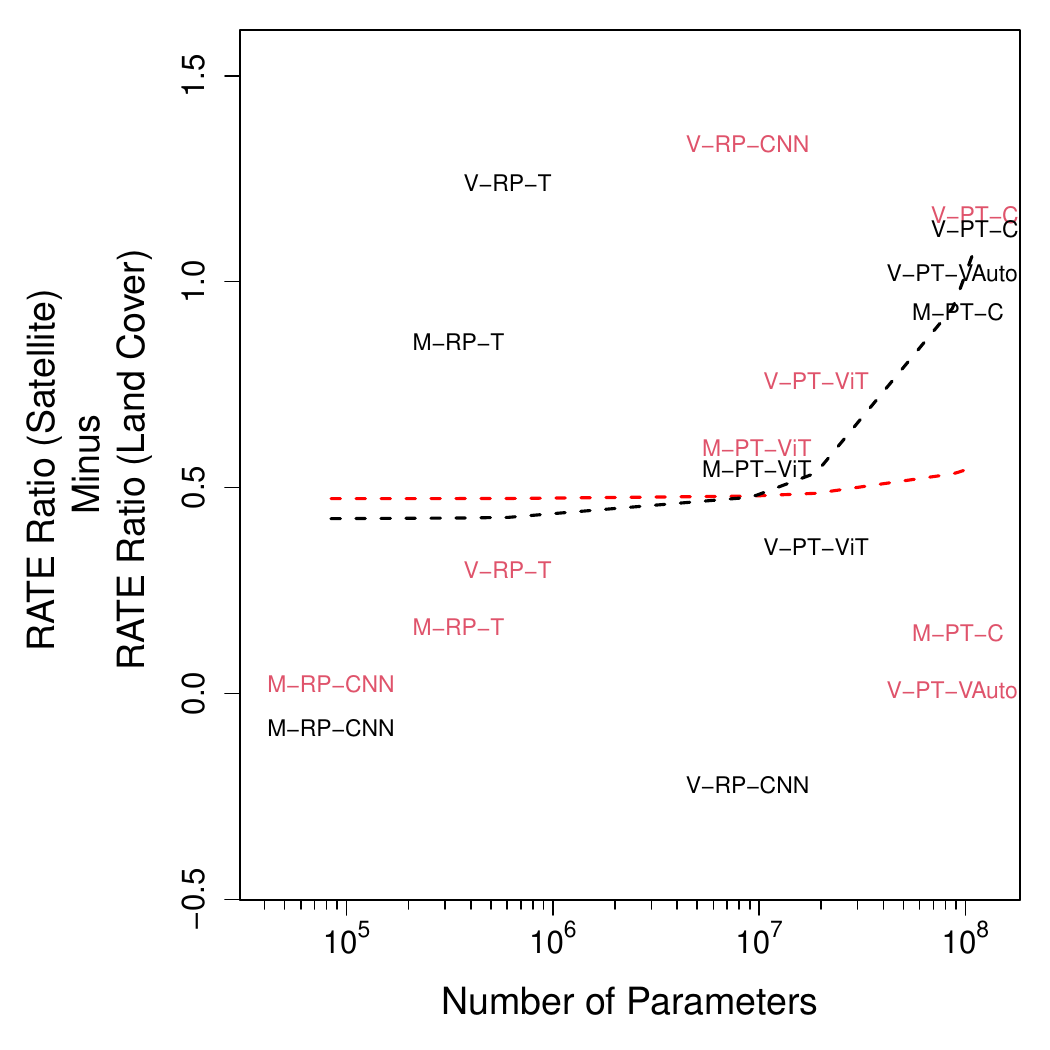}
\caption{
Lift from use of raw satellite imagery, as quantified by difference between the RATE measure from satellite analysis and from land cover analysis (positive values indicate relative benefit of raw EO representation over land cover representation). \RATEWTALT{} weighting used in RATE analysis. Dashed lines indicating local averages are computed using splines; dotted red/black lines indicate RATE ratios for the tabular-only baseline. 
}
\label{fig:LandcoverRATE\RATEWTALT}
\end{center}
\end{figure}

\begin{figure}[ht!] 
  \begin{center}
\includegraphics[width=0.45\linewidth]{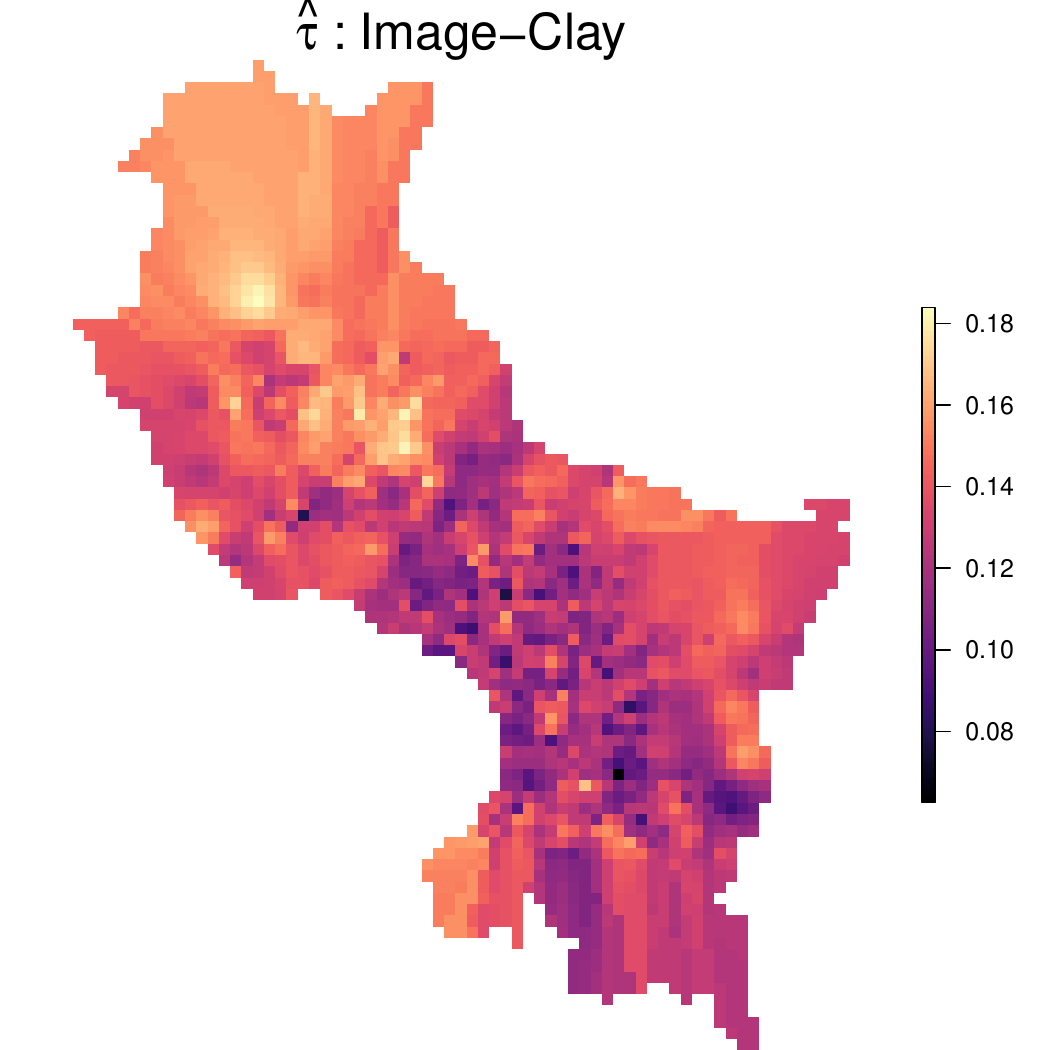}
\includegraphics[width=0.45\linewidth]{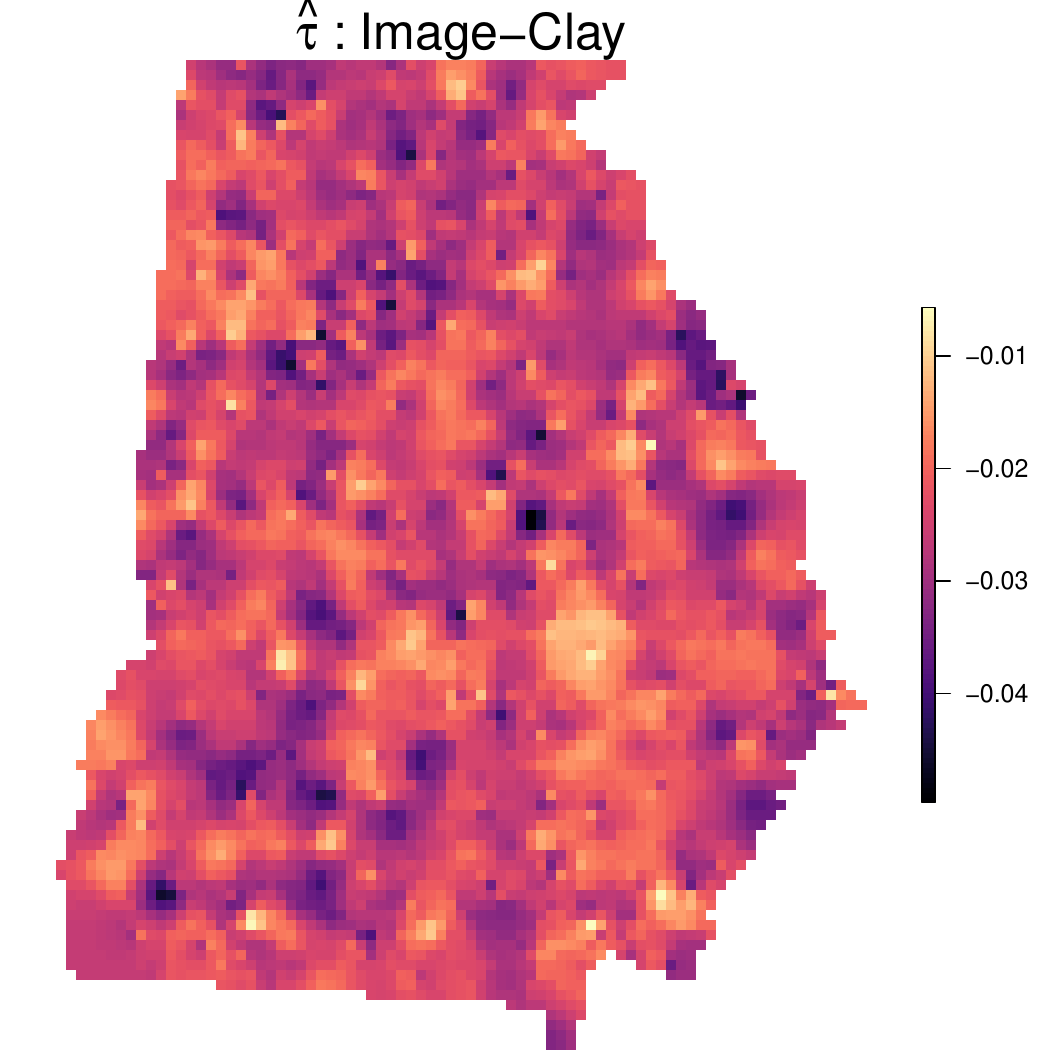}
\caption{Transportability analysis using image sequence data in Cusco {\sc (Left)} and Georgia {\sc (Right)}. Colors indicate treatment effect estimate values under the model specification  (yellow $>$ blue). Clay model with images used.}
\label{fig:TransportClayImage}
\end{center}
\end{figure}

\begin{figure}[ht!] 
  \begin{center}
\includegraphics[width=0.45\linewidth]{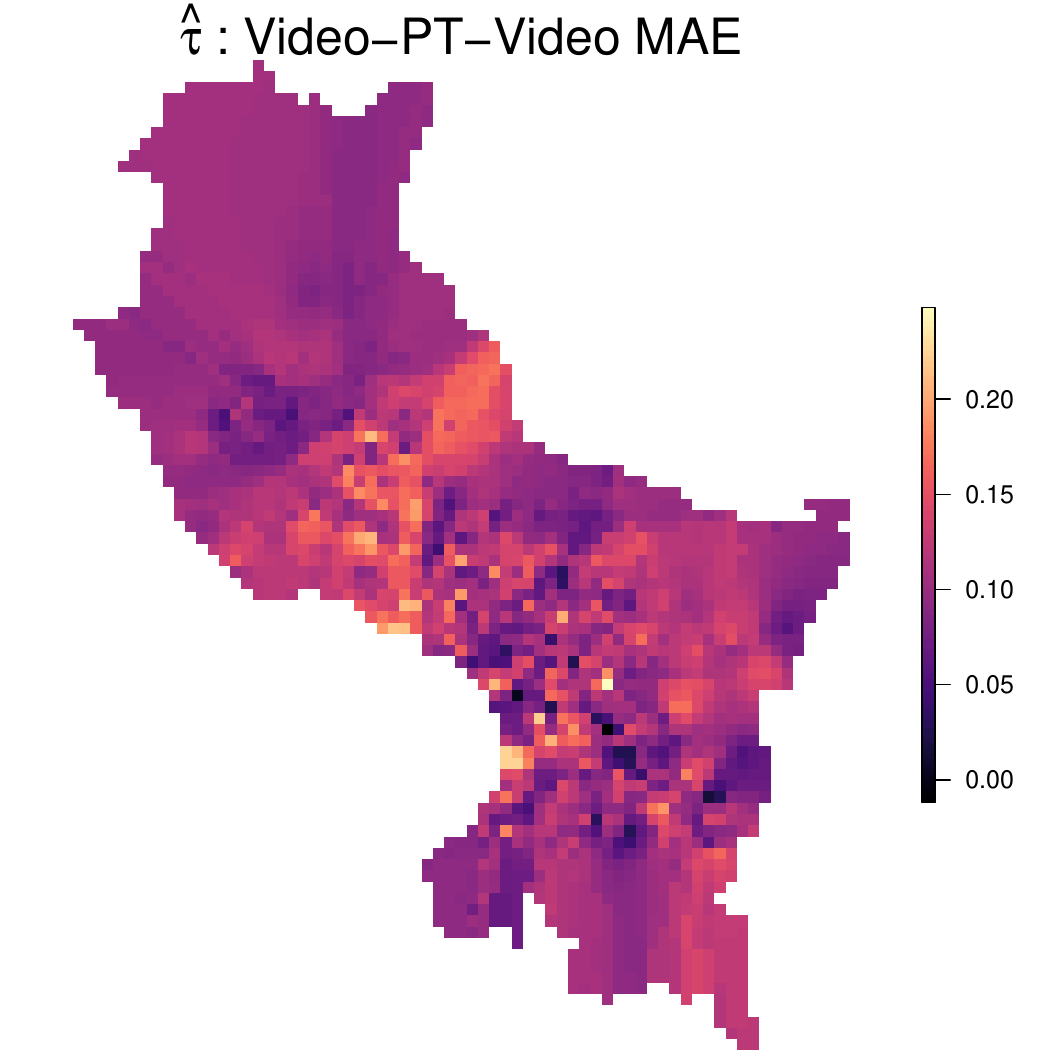}
\includegraphics[width=0.45\linewidth]{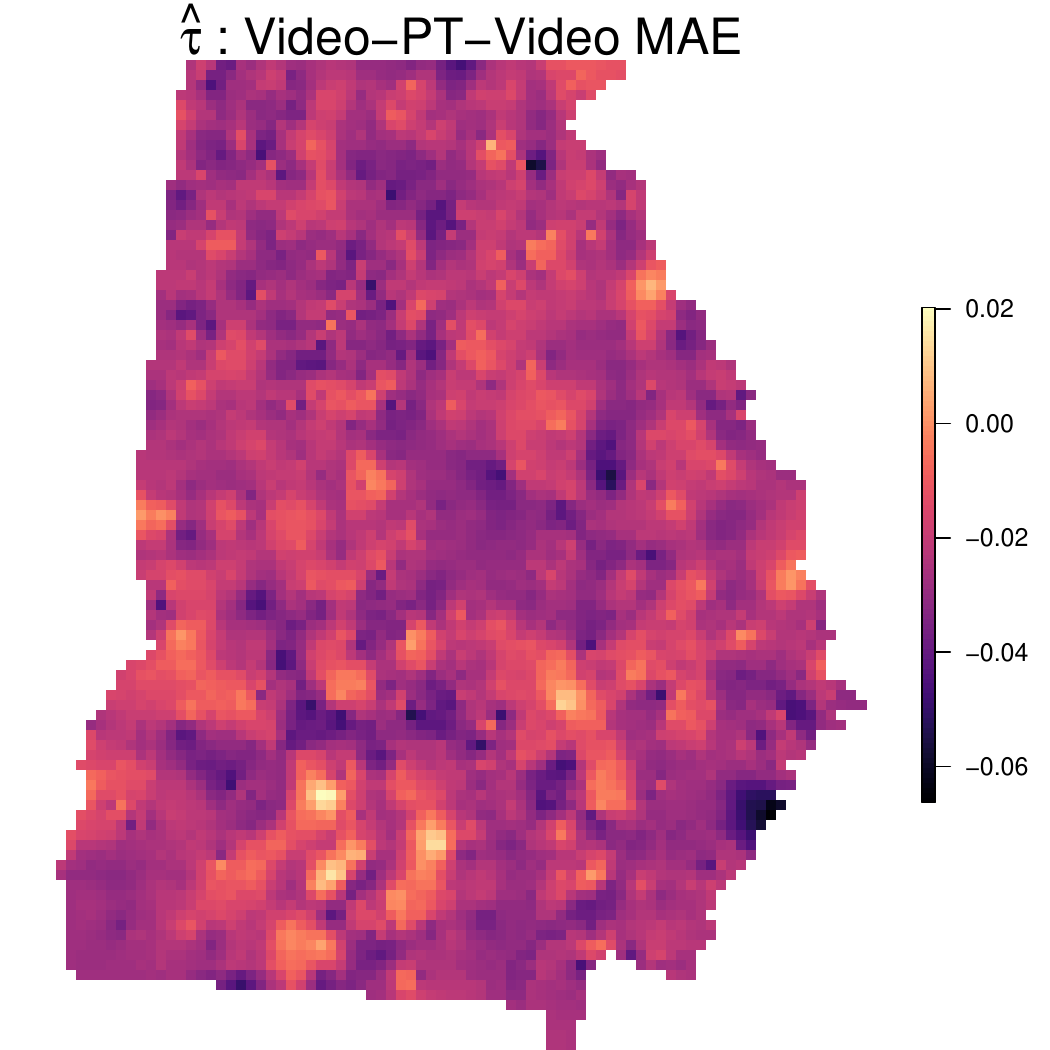}
\caption{Transportability analysis using image sequence data in Cusco {\sc (Left)} and Georgia {\sc (Right)}. Colors indicate treatment effect estimate values under the model specification  (yellow $>$ blue). Masked video autoencoder in image sequence representation.}
\label{fig:TransportVideoMAE}
\end{center}
\end{figure}

\clearpage\newpage 
\renewcommand{\thefigure}{A.III.\arabic{figure}}
\setcounter{figure}{0}  

\renewcommand{\thetable}{A.III.\arabic{table}}
\setcounter{table}{0}  

\renewcommand{\thesection}{A.III.\arabic{section}}
\setcounter{section}{1}
\section*{Appendix III. Generating Land Cover Summaries}

In the main text, we explore the relative heterogeneity content of EO image sequences and land cover maps, which are usually generated from satellite imagery and form a discrete classification over pixels (usually at a 30 m by 30 m resolution). Classifications can take on environmental values such as ``Swamp'' and ``Woodland'', or values related to human use (such as ``Mines'' and ``Urban'').

We obtain pre-treatment Georgia land cover data from the USGS GAP/LANDFIRE National Terrestrial Ecosystem Survey. We obtain pre-treatment Peru land cover data from the MapBiomas
initiative. 

For each location in Georgia and Peru, we summarize the location of interest using the land cover maps as follows. For the same spatial domain around each point, we calculate the proportion of land cover classes in the area. In essence, we summarize what proportion of each raw image is classified in the various land cover categories. To address the fact that many of the land cover classes have proportions near 0, we take the logit transformation, so that the proportions now line on the real line for use in the downstream heterogeneity models. We then apply the effect heterogeneity models to these land cover representations.

\clearpage\newpage 
\renewcommand{\thefigure}{A.IV.\arabic{figure}}
\setcounter{figure}{0}  

\renewcommand{\thetable}{A.IV.\arabic{table}}
\setcounter{table}{0}  

\renewcommand{\thesection}{A.IV.\arabic{section}}
\setcounter{section}{1}

\section*{Appendix IV.  Glossary}
The following abbreviations are used in this paper: 
\begin{itemize}
\item {\bf ATE}: Average Treatment Effect. The mean difference between the outcomes under treatment and control conditions for the entire population.
\item {\bf AUTOC}: Area Under the Treatment Effect Curve. A weighting mechanism used in the RATE calculation that assesses the cumulative gain in treatment effect lift over the ATE as more individuals are treated, sorted by their predicted treatment effect.
\item {\bf CATE}: Conditional Average Treatment Effect. The average treatment effect for a subpopulation defined by specific pre-treatment characteristics.
\item {\bf CNN}: Convolutional Neural Network. A type of deep learning model commonly used for analyzing visual imagery.
\item {\bf EO}: Earth Observation. The gathering of information about Earth's physical, chemical, and biological systems via remote sensing technologies, typically satellites.
\item {\bf PC}: Principal Component. A technique used for dimensionality reduction based on maximizing the variability present in the projected orthogonal directions.
\item {\bf QINI}: A weighting mechanism used in the RATE calculation that measures the area between the QINI curve (which plots the cumulative treatment effect against the proportion of the population treated) and the 45-degree line.
\item {\bf R-learner}: A method for estimating heterogeneous treatment effects. 
\item {\bf RATE}: Rank-Weighted Average Treatment Effect. A measure used to quantify and compare the performance of different CATE estimation methods in the absence of ground truth individual treatment effects.
\item {\bf RCT}: Randomized Controlled Trial. An experimental design where participants are randomly assigned to either a treatment or control group to test the efficacy of an intervention.
\item {\bf ViT}: Vision Transformer. A type of neural network architecture designed for image analysis tasks, inspired by Transformers from natural language processing (NLP) and using an attention matrix capturing interdependencies between all segments of the image with all others. 
\end{itemize}

\clearpage
\newpage 
\renewcommand{\thesection}{A.V.\arabic{section}}
\setcounter{section}{1}

\section*{Appendix V. Computer Vision Models for EO Heterogeneity Analysis}

This Appendix provides a more detailed description of the computer vision models used in our analysis. These models are employed to generate representations from image sequences, which are then used in our causal inference framework.

\subsection{Convolutional Neural Networks (CNNs)}

Convolutional Neural Networks are a class of deep learning models effective for analyzing visual imagery. They are designed to characterize information from input images, such as texture. Key CNN components: 
\begin{itemize}
    \item \textbf{Convolutional layers:} These layers apply a set of learnable filters to the input, creating feature maps that highlight important aspects of the image.
    \item \textbf{Pooling layers:} These reduce the spatial dimensions of the feature maps, helping to achieve spatial invariance.
    \item \textbf{Fully connected layers:} These layers typically appear at the end of the network and produce the final output.
\end{itemize}

In our analysis (i.e., in the randomized projection approach using a CNN for spatial aggregation), we apply a randomly initialized single-layer CNN, followed by global maximum pooling across the image array (followed by a randomized Transformer for temporal aggregation).

\subsection{Vision Transformers (ViT)}

Vision Transformers (ViTs) adapt the Transformer architecture, originally designed for natural language processing, to image analysis tasks. The key idea is to treat an image as a sequence of patches and process these patches using self-attention mechanisms.

Key components of ViTs include:

\begin{itemize}
    \item \textbf{Patch embedding:} The input image is divided into fixed-size patches, which are then linearly embedded.
    \item \textbf{Position embedding:} This is added to provide information about the relative or absolute position of the patches.
    \item \textbf{Transformer encoder:} This consists of alternating layers of multiheaded self-attention and Multi-layer Perception (i.e. Feed-forward Neural Network) blocks. Self-attention layers model the relationship between each image patch with every other image patch. The self-attention modules can be seen as a learnable fully connected graph between all image patches. 
\end{itemize}
We use both a randomized ViT and a pre-trained ViT in our analysis for spatial aggregation. Again, a randomized temporal ViT aggregates across time.

\subsection{Video Masked Autoencoder}

The Video Masked Autoencoder is an unsupervised learning approach for video understanding. Key aspects:

\begin{itemize}
    \item \textbf{Masking:} A portion of the input video frames or patches are masked (i.e., held out as the target of prediction).
    \item \textbf{Encoder:} This module processes the visible portions of the video into a vector representation (used as the representation for other pre-trained tasks, as here). 
    \item \textbf{Decoder:} This model takes the output of the encoder and attempts to reconstruct the masked image portions. 
\end{itemize}
The model learns to understand video content by trying to predict the masked portions based on the visible parts, capturing both spatial and temporal dependencies.

\subsection{Clay Foundation Model}

The Clay Foundation Model is specifically designed for Earth Observation (EO) tasks. It is pre-trained on a large corpus of satellite imagery data from various sources including Landsat and Sentinel satellites. Again, a portion of the input EO images are masked (i.e., held out as the target of prediction), with an encoder-decoder architecture then used. We take the output of the encoder model as the satellite image representation. See \href{https://madewithclay.org/}{\url{MadeWithClay.org}} for more information.

\end{document}